\documentclass[aps,pra,reprint,amsmath,amssymb,superscriptaddress,onecolumn,nofootinbib,notitlepage]{revtex4-2}
\usepackage{mathtools}
\usepackage{siunitx}
\usepackage[hidelinks]{hyperref}
\usepackage{svg}
\usepackage{caption}
\usepackage{float}
\usepackage{subcaption}
\usepackage[braket, qm]{qcircuit}
\usepackage{bbm}
\usepackage{lipsum} 
\usepackage{comment}
\usepackage{bm} 
\usepackage{float}

\begin{document}
\title{Large-scale quantum reservoir computing using a Gaussian Boson Sampler}

\author{Valeria~Cimini}
\email{To whom correspondence should be addressed: valeria.cimini@uniroma1.it, pmcmahon@cornell.edu}
\affiliation{Dipartimento di Fisica, Sapienza Università di Roma, P.le Aldo Moro 5, I-00185 Roma, Italy}
\affiliation{School of Applied and Engineering Physics, Cornell University, Ithaca, NY 14853, USA}

\author{Mandar~M.~Sohoni}
\affiliation{School of Applied and Engineering Physics, Cornell University, Ithaca, NY 14853, USA}

\author{Federico~Presutti}
\thanks{Present address: Research Laboratory of Electronics, Massachusetts Institute of Technology, Cambridge, MA 02139, USA}
\affiliation{School of Applied and Engineering Physics, Cornell University, Ithaca, NY 14853, USA}

\author{Benjamin~K.~Malia}
\thanks{Present address: Air Force Research Laboratory, Rome, NY 13441, USA}
\affiliation{School of Applied and Engineering Physics, Cornell University, Ithaca, NY 14853, USA}

\author{Shi-Yuan~Ma}
\thanks{Present address: Research Laboratory of Electronics, Massachusetts Institute of Technology, Cambridge, MA 02139, USA}
\affiliation{School of Applied and Engineering Physics, Cornell University, Ithaca, NY 14853, USA}

\author{Ryotatsu~Yanagimoto}
\affiliation{School of Applied and Engineering Physics, Cornell University, Ithaca, NY 14853, USA}
\affiliation{Physics \& Informatics Laboratories, NTT Research, Inc., Sunnyvale, CA 94085, USA}

\author{Tianyu~Wang}
\affiliation{School of Applied and Engineering Physics, Cornell University, Ithaca, NY 14853, USA}

\author{Tatsuhiro~Onodera}
\affiliation{School of Applied and Engineering Physics, Cornell University, Ithaca, NY 14853, USA}
\affiliation{Physics \& Informatics Laboratories, NTT Research, Inc., Sunnyvale, CA 94085, USA}

\author{Logan~G.~Wright}
\thanks{Present address: Department of Applied Physics, Yale University, New Haven, CT 06520, USA}
\affiliation{School of Applied and Engineering Physics, Cornell University, Ithaca, NY 14853, USA}
\affiliation{Physics \& Informatics Laboratories, NTT Research, Inc., Sunnyvale, CA 94085, USA}

\author{Peter~L.~McMahon}
\email{To whom correspondence should be addressed: valeria.cimini@uniroma1.it, pmcmahon@cornell.edu}
\affiliation{School of Applied and Engineering Physics, Cornell University, Ithaca, NY 14853, USA}
\affiliation{Kavli Institute at Cornell for Nanoscale Science, Cornell University, Ithaca, NY 14853, USA}

\begin{abstract}

A Gaussian boson sampler (GBS) is a special-purpose quantum computer that can be practically realized at large scale in optics because it involves only squeezing, beamsplitter unitaries, and photon-counting measurements. While Gaussian Boson Samplers have been used to demonstrate quantum computational supremacy, an open challenge is to discover what practical tasks a GBS could effectively be applied to. Here we report on experiments in which we used a frequency-multiplexed GBS with $>400$ modes as the reservoir in the quantum-machine-learning approach of quantum reservoir computing. We evaluated the accuracy of our GBS-based reservoir computer on a variety of benchmark tasks, including spoken-vowels classification and MNIST handwritten-digit classification. We found that when the reservoir computer was given access to the correlations between measured modes of the GBS, the achieved accuracies were the same or higher than when it was only given access to the mean photon number in each mode---and in several cases the advantage in accuracy from using the correlations was greater than 20 percentage points. This provides experimental evidence in support of theoretical predictions that access to correlations enhances the power of quantum reservoir computers. We also tested our reservoir computer when operating the reservoir with various sources of classical rather than squeezed (quantum) light: coherent laser light, thermal light from amplified spontaneous emission, and supercontinuum light generated using a highly nonlinear fiber. We found that using squeezed light consistently resulted in the highest (or tied highest, for simple tasks) accuracies. Our work experimentally establishes that a GBS can be an effective reservoir for quantum reservoir computing and shows the possibility of experimentally exploring the power of GBS-based quantum reservoir computing, the role of quantumness and correlations in large-scale quantum reservoir computing, and the applications of quantum reservoir computing.
\end{abstract}

\maketitle

\section{Introduction}
\label{sec:intro}

Quantum photonics has recently been used to experimentally demonstrate quantum computational advantage \cite{zhong2020quantum,madsen2022quantum} with Gaussian boson sampling (GBS) \cite{hamilton2017gaussian}.
GBS is based on sampling multimode squeezed states, which in the optical domain are a scalable and useful resource for quantum networks \cite{roslund2014wavelength} and for cluster-state-based quantum computing \cite{menicucci2006universal,briegel2009measurement,raussendorf2007topological,yokoyama2013ultra,asavanant2019generation,larsen2019deterministic}. 
The demonstrations of quantum computational advantage with GBS \cite{zhong2020quantum,madsen2022quantum} have been on sampling tasks that are designed to be difficult for classical processors to solve but that are mathematically abstract and not directly connected with any practical applications. Finding practical uses for GBS that simultaneously deliver better-than-classical solutions while also being hard to simulate classically---thereby yielding a practical quantum advantage---has proven extremely challenging. For example, vibronic-spectra simulation has been a leading candidate application for GBS for a decade \cite{huh2015boson}, with strong intuitive reasons to expect that GBS would have an advantage at the task in part because of the close match between what GBS naturally does and the task being solved, and yet a large fraction of use cases have recently been dequantized \cite{oh2024quantum}.\footnote{This is in the setting of an ideal---lossless---GBS apparatus. When one considers the experimentally relevant setting of GBS with photon losses, the challenge to avoid classical simulability is even harder \cite{oh2024classical}.} In this paper, we will not report a full solution to this grand challenge of finding a practical application for which GBS provides a quantum advantage (over all possible classical methods), but instead address part of the challenge: exploring an application for which GBS can at least perform the desired tasks well, and experimentally provides an advantage over a classicalized version of our experimental apparatus.

Quantum machine learning (QML) \cite{biamonte2017quantum}---the use of quantum computers to perform machine learning---has attracted much attention since the advent of the era of noisy, intermediate-scale quantum (NISQ) machines \cite{preskill2018quantum}. However, initial hopes that NISQ machines could deliver quantum advantage for practical machine-learning problems on classical data have been dampened by a range of challenges \cite{cerezo2022challenges}. Even getting NISQ machines to deliver accurate machine-learning predictions without worrying about whether they are doing so in a way that gives any advantage over classical methods has proven very difficult, in large part because of the difficulty of loading the data to be classified into the quantum computer reliably in the first place. For example, Ref.~\cite{johri2021nearest} reported a prediction accuracy of ${\sim}$78\% on MNIST handwritten-digit classification using a commercial trapped-ion quantum processor, while the classical state-of-the-art for the same task from 1998 was ${\sim}$99\% \cite{lecun1998gradient}.\footnote{Recently Ref.~\cite{kornjavca2024large} reported a similar QML solution and demonstration to ours for MNIST handwritten-digit classification, achieving similar results, as we discuss at the end of our paper.}

A variety of QML algorithms have been developed, but the vast majority of NISQ demonstrations have been performed with variational quantum circuits \cite{cerezo2021variational} in a hybrid quantum--classical approach that realizes either a quantum neural network or a quantum kernel method \cite{cerezo2022challenges}. In variational quantum algorithms, a quantum computer runs a quantum circuit whose gates are parameterized, and the expectation values of various observables are used as inputs to a classical calculation of an objective function (loss function, in the language of machine learning---where the loss may characterize how accurately the machine-learning system is making predictions across the set of training samples). A classical optimizer is used to update the variational parameters in a way that ideally minimizes the objective function, and this process of estimating the function using the quantum computer and then updating the parameters to try minimize the function is then repeated in a loop. Unfortunately, in addition to challenges in loading classical data, these variational approaches have often proven difficult to optimize (i.e., train, in the language of machine learning) when scaled up due, in part, to the issue of barren plateaus in the landscape of gradients \cite{mcclean2018barren}, rendering gradient-based optimization ineffective \cite{cerezo2022challenges}. A class of QML algorithms suitable for NISQ processors that avoids the optimization challenges of variational quantum algorithms is quantum reservoir computing (QRC) \cite{fujii2017harnessing,fujii2021quantum}. In QRC, an untrained quantum system is used to process input data in a way that---if the quantum system is chosen well---ideally generates useful\footnote{Here \textit{useful} means that the computed features extract the relevant information from the input data for the prediction/classification task at hand, such that the subsequent classical linear classifier can make accurate predictions in spite of the fact that it only performs linear processing.} high-dimensional, nonlinear features of the input that are then fed into a single classical linear classifier, which can make a prediction (classification) based on those features. The classical linear classifier is trained---but a key benefit of QRC (versus, for example, variational quantum approaches) is that training a classical classifier is far easier than training a quantum system, needing far fewer executions of the quantum system (reservoir) and not suffering from barren plateaus \cite{fujii2021quantum}. When acting on classical input data, if the features computed by the quantum reservoir are difficult to compute classically, QRC may in principle provide a quantum computational advantage. However, a useful quantum advantage can only result if the QRC is capable of achieving better performance on one or more practical machine-learning tasks than all alternative classical machine-learning methods---which is both hard to achieve, and hard to prove.\footnote{As we mentioned earlier in the introduction, we don't address the question of how to achieve quantum advantage in the experiments reported in this paper, but focus instead on how well the reservoir-computer approach can perform on different tasks, under different experimental conditions, and with different choices of generated features.}
QRC has recently been explored both theoretically and experimentally in photonic systems \cite{Wrightpreprint,garcia2023scalable,suprano2024experimental,garcia2024squeezing,nerenberg2024photon,zia2025quantum,spagnolo2022experimental,selimovic2025experimental}, as well as with superconducting circuits \cite{yasuda2023quantum,senanian2024microwave,hu2024overcoming} and Rydberg atoms \cite{bravo2022quantum,kornjavca2024large}. QML protocols besides QRC have also been investigated theoretically and experimentally in photonics \cite{steinbrecher2019quantum,ewaniuk2023imperfect,yin2024experimental,hoch2025quantum}.

Continuous-variable (CV) quantum systems, which can naturally be realized in optics \cite{fukui2022building}, have a potential hardware-efficiency benefit for QML \cite{killoran2019continuous,garcia2024quantum}: a single CV quantum mode can encode and process more information than is possible with a single qubit due to the use of a higher-dimensional Hilbert space\footnote{Sometimes CV systems are described as having a benefit from using an infinite-dimensional Hilbert space versus the finite-dimensional Hilbert space of qubits \cite{fukui2022building}, but in the setting of bosonic modes under the practical constraint of using finite energy---imposing a finite maximum number of excitations of each mode---the benefit is probably better thought of as enabling a constant-factor improvement in the efficient use of hardware resources.}. As a result, CV quantum-optical systems have been proposed for a variety of QML, including the implementation of quantum kernel methods \cite{anai2024continuous}, and as quantum reservoirs \cite{Wrightpreprint,nokkala2021gaussian,mujal2021opportunities}

In this paper, we report on experiments we have conducted with an apparatus that is a frequency-domain Gaussian boson sampler, which we use as a quantum reservoir for QRC.\footnote{We use the term \textit{quantum reservoir computing} (QRC) for any scheme in which data is input to a quantum system and then results of measurements of that quantum system are sent to a trained classical linear neural-network layer to make a prediction. When the input data are not time series and/or the quantum system doesn't feature some explicit recurrence (such as by passing processed measurement results back as inputs into the system), the name \textit{quantum extreme learning machine} (QELM) is sometimes used in the literature \cite{mujal2021opportunities,xiong2023fundamental}. Since in this paper we report experiments that are only on time-independent data, one could also reasonably describe our work as demonstrating a QELM. We have used the more general term QRC because nothing in our system or scheme prevents them from, in principle, being used on time-series inputs too, and a QELM is sometimes also called a \textit{feedforward QRC}.} The experimental system acting as the quantum reservoir is described in detail in Ref.~\cite{presutti2024highly}; it comprises three stages, which we will explain further in the next section: (1) \textit{generation} of $>400$ modes of squeezed light; (2) \textit{frequency-domain-beamsplitter unitary} evolution, where the beamsplitter unitary is programmable---it is through this programmability that we input classical data to the reservoir to process; (3) \textit{measurement} of the state by frequency-resolved photodetection---we use the measurement results to estimate the mean number of photons per mode and two-body correlations between modes, which we send as outputs from the reservoir to the classical linear layer of the QRC. We tested QRC using this quantum reservoir with several multiclass classification tasks each based on classical data: two tasks that are standard machine-learning tests based on synthetic data for nonlinear classifiers and two tasks that are based on real-world data (spoken-vowel and handwritten-digit classification).

\clearpage

We found that the quantum reservoir computer was able to achieve accuracies substantially surpassing that of a conventional, purely linear classifier on the synthetic-data tasks, and also achieved high ($>97\%$) accuracy on the vowel-classification task. Across all tasks, our results showed that using the measured two-body correlations resulted in substantially improved accuracies versus just giving the QRC classical linear layer access to the mean number of photons per mode. In addition to using our experimental system as a QRC, we also repeated our experiments on the synthetic-data tasks with three different choices of classical input light instead of multimode squeezed light (quantum light): coherent light (from a continuous-wave laser), thermal light (from amplified spontaneous emission (ASE)), and supercontinuum light (from a nonlinear-fiber supercontinuum source)---rendering the system a classical optical reservoir computer. We found that the accuracies achieved when using the squeezed light were substantially higher than when using the thermal or supercontinuum light, and were consistently, albeit within the ranges of statistical uncertainty, higher than when using the coherent laser light. We also repeated our experiments on vowel classification with coherent laser light, since this was the most competitive with squeezed light in the synthetic tasks---and we again found that using squeezed light consistently resulted in better accuracies than using coherent light, albeit again also within the statistical uncertainty\footnote{Although within statistical error bounds, the gaps in accuracies between the squeezed- and coherent-light cases in Fig.~4a and Fig.~4b appear so consistently that it seems more reasonable to hypothesize that squeezed light truly does result in better accuracies, than to hypothesize that there is no difference. Resolving the question (of whether and when squeezed light gives superior accuracies) with greater statistical certainty remains for future work though.}. In summary, in this paper we show how by using a quantum-optical apparatus with $>400$ modes, and $>100$ programmable degrees of freedom with which one can input classical data to be classified, we have been able to successfully perform QRC on classical data with very high dimensionality (feature vector dimension of up to $100$ in the case of the MNIST handwritten-digit classification task).

\section{Classification protocol and experimental setup}
\label{sec:protocol}

In the experiments we report in this paper, we followed the standard approach to training and testing a quantum reservoir computer (QRC) for time-independent classification tasks. This section is organized as follows: for clarity we first summarize the generic procedure for training and operating a reservoir computer; we then explain our physical reservoir at a high level, how it can be operated as a quantum or a classical reservoir, and how the measurements of the output of the reservoir are used; we then describe the experimental setup realizing our reservoir and explain how data to be classified is input to the reservoir; finally, we explain some important details about how we switch between quantum and classical modes of operation.

Reservoir computing (RC) \cite{lukovsevivcius2009reservoir} is a class of supervised models in which data is sent through a fixed (untrained) neural network (the reservoir) and then to a trained linear output layer. Physical systems can also act as reservoirs \cite{tanaka2019recent}, and in QRC, the physical system is a quantum system. In this work, we study both QRC and several variants of classical physical RC, keeping each system as much as possible on the same playing field when switching between quantum and classical variants. In all these variants of reservoir computing, a key benefit over other quantum or classical machine-learning approaches is that training the reservoir computer to perform classification (or other tasks---but in this work we focus on classification tasks) involves training only the linear output layer, which maps measurements of the reservoir state to a final output (classification result). Suppose we have a training data set $\{(\mathbf{x}^{(k)},l^{(k)})\}$ where $\mathbf{x}^{(k)}$ is the $k$th example of input data (sometimes called the \textit{feature vector} because it describes the features of whatever is being classified), and $l^{(k)}$ is the corresponding label for that example. During training, examples of input data are sent into the reservoir and the output of the reservoir is measured. In the setting of time-independent data, for each input example, the reservoir effectively performs a (potentially nonlinear) function on the input vector $\mathbf{x}$, outputting a vector that summarizes the measurement results: $f(\mathbf{x})$. The linear output layer multiplies the output vector from the reservoir by a matrix $W$, yielding a final output vector $\mathbf{y} = W f(\mathbf{x})$. Typically the elements of $\mathbf{y}$ are treated as corresponding to the possible labels that the data can be classified as, and the index of the largest element of $\mathbf{y}$ is reported as the classification result. The matrix $W$ is trained by choosing its elements so that the classification results are correct for as many of the examples in the training data set as possible. In this work, we demonstrate that achieved performance with our chosen quantum reservoir is robust to the training method $W$: we compared the accuracies achieved when using the pseudoinverse method \cite{lukovsevivcius2009reservoir}, stochastic gradient descent, logistic regression, and a support vector machine (SVM) with a linear kernel (see Subsection~\ref{Generic_subsec}). Since the accuracies were very similar for all the training methods, we used the SVM method for all the other classification tasks (see Appendix \ref{App_opt} for details).

\begin{figure*}[htb!]
    \includegraphics[width=0.99\textwidth]{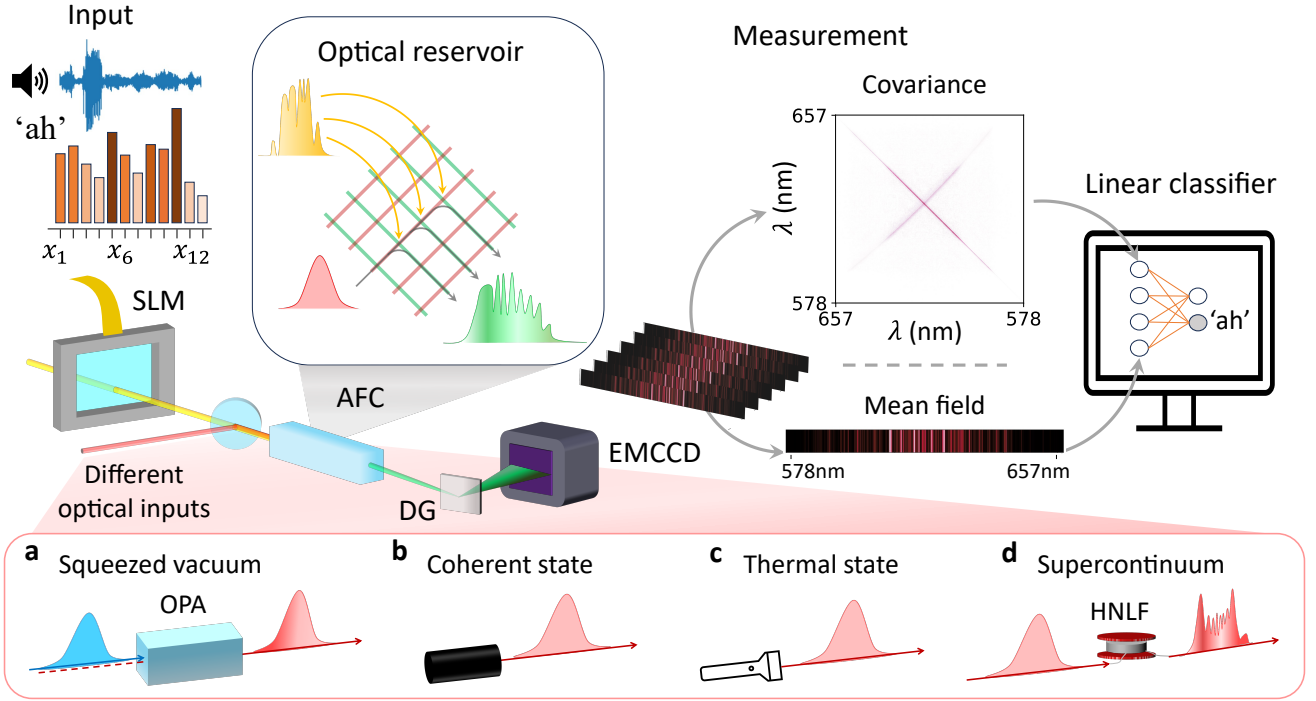}
    \caption{\textbf{Quantum and classical reservoir computing for classification of classical data using a frequency-domain optical reservoir with $>400$ frequency modes.} The features of the classical sample to classify, represented by a vector $\mathbf{x}$, are optically encoded by shaping the spectral phases $\phi(\lambda)$ of a classical pump (yellow) of an adiabatic frequency conversion (AFC) process by using a spatial light modulator (SLM) in a pulse shaper. The AFC process acts as a programmable frequency-domain beamsplitter unitary on the signal input (red), where what unitary is performed is determined by $\phi(\lambda)$ \cite{presutti2024highly}. The states of the frequency modes of the converted beam (green) are therefore affected by the input data $\mathbf{x}$. The frequency modes are measured using an EMCCD camera after sending the converted beam through a diffraction grating (DG). Each camera frame effectively contains a single-shot measurement of the frequency modes in the photon-number (Fock) basis. For a single input $\mathbf{x}$, multiple camera frames are taken, from which we construct either a photon-number covariance matrix (flattened to form a vector) or a mean-field vector containing the average photon number in each mode; either of these is then treated as the output vector of the optical reservoir, $f(\mathbf{x})$. The outcome vectors are then fed to a digital linear classifier trained to classify the corresponding input. We studied the achieved classification accuracy when different weak (low-average-photon-number) optical states were used as the signal inputs (red): a) squeezed vacuum generated by a pulsed optical parametric amplifier (OPA), b) coherent states generated by a $1550\,\textrm{nm}$ CW laser, c) thermal light emitted from an erbium-doped fiber amplifier (EDFA), and d) broadband supercontinuum light generated using a highly nonlinear fiber (HNLF).}
	\label{fig1}
\end{figure*}

We study physical RC using an \textit{optical} reservoir that acts as a quantum reservoir when given a quantum state as a resource input, and acts as a classical reservoir if the resource input is a classical state of light (Fig.~\ref{fig1}). In addition to taking in a resource input that can be quantum or classical, our physical reservoir also takes in another purely classical state of light that encodes the input data to be classified. The output of the reservoir computation is determined by measuring the optical state emitted from the reservoir using an electron multiplying charge-coupled device (EMCCD) camera that effectively acts as a one-dimensional array of single-photon-sensitive detectors. To measure the output of the reservoir $f(\mathbf{x})$ for a sample $\mathbf{x}$ from the dataset, the reservoir computation is repeated multiple times: each time, the same sample is input to the reservoir and the output light from the reservoir is measured with the EMCCD camera, resulting in a stored camera frame. In postprocessing on a digital computer, we compute a covariance matrix (described formally later in this section) using the multiple camera frames, and it is this covariance matrix---which contains information about the correlations between camera pixel measurements, and hence the optical modes---that is used as the reservoir output vector\footnote{We \textit{flatten} the $M \times M$ covariance matrix to turn it into an $M^2$-dimensional vector.} $f(\mathbf{x})$. Finally, this vector is passed through a trained linear classifier, computing $\mathbf{y} = W f(\mathbf{x})$, which gives the classification result. In our experiments, we compare the above procedure with an alternative in which the camera frames recording the reservoir output are post-processed differently: instead of computing a covariance matrix, we simply compute the mean photon number of each optical mode and use this mean-field vector as $f(\mathbf{x})$. With this procedure, information about the correlations between optical modes is discarded, so we are able to experimentally study the importance of correlation measurements in the physical reservoirs we tested.

The core physical process in our optical reservoir is adiabatic frequency conversion (AFC) \cite{moses2012fully}, which is a nonlinear-optical process that converts light from one center wavelength to another.\footnote{AFC---as opposed to regular nonlinear-optical frequency conversion---enables broadband light to be converted with high quantum efficiency, which is a technical feature that we take advantage of in our experiments and is explained in detail in Ref.~\cite{presutti2024highly}.} Importantly, from the perspective of what computation the reservoir performs, AFC can be thought of as applying a unitary beamsplitter transformation of input modes to output modes\footnote{The output modes happen to be at different wavelengths than the input modes; this is unimportant from the perspective of the computation that is performed (it is just a unitary mode transformation), but important from a technological perspective (it allows input modes at wavelengths (such as 1550 nm) at which it is easy to produce squeezed light but challenging to perform photodetection to be converted to output modes at visible wavelengths, for which sensitive and relatively inexpensive (per pixel) detector arrays are commercially available).}. An optical input state with a central wavelength near $\SI{1550}{\nano\meter}$ is up-converted in frequency using a pp-potassium--titanyl--phosphate (pp-KTP) crystal pumped by a broadband pulsed laser with a central wavelength at $\SI{1033}{\nano\meter}$. The data (feature vector) for a sample to classify, $\mathbf{x}$, is input into the optical reservoir by modulating the phases of the AFC pump pulse as a function of wavelength, $\phi(\lambda)$, using a spatial light modulator (SLM) in a grating-based pulse shaper. The pulse shaper's SLM has 1920 pixels in a single row, allowing the phases $\phi(\lambda)$ for at most 1920 different wavelengths\footnote{Technically ranges, or bins, of wavelength.} to be programmed. The feature vector $\mathbf{x}$ is programmed into these 1920 pixels using a straightforward mapping of $\mathbf{x}$ to pixels that is not tailored to the task---this results in the phase for the $i$th discrete bin of wavelength, $\phi(\lambda_i)$, depending on the $i$th element of the input vector, $\mathbf{x}_i$ (see Methods for details). The number of pixels, 1920, gives an upper bound on the size of a single vector $\mathbf{x}$ that can be input to the reservoir in a single shot, but in practice the effective number of wavelengths we could independently control the phases of in this study was approximately 100 (see Methods), and so the highest-dimensional dataset we report experiments with is MNIST with 100-dimensional vectors $\mathbf{x}$.

The phases $\phi(\lambda)$ imprinted on the pump pulse by the pulse shaper directly influence the nonlinear dynamics of the optical reservoir. One can think of them as controlling the unitary transformation that the AFC process performs on the weak input (signal) state, resulting in nonlinear processing of the data in a similar way to recent work showing that inputting data into classical linear-optical neural networks via the parameters of the system (as opposed to via the optical inputs to the system) gives rise to nonlinear computation \cite{wanjura2024fully,xia2024nonlinear,yildirim2024nonlinear,mcmahon2024nonlinear}. The fact that data encoded in unitaries in quantum systems can undergo nonlinear transformations has also been explained in the quantum-machine-learning literature, for example in Refs.~\cite{mitarai2018quantum,schuld2019quantum,Wrightpreprint}. When the input state is squeezed, one can also view the AFC unitary as being responsible for generating complex patterns of correlation in the output frequency modes that depend on the phases of the AFC pump, i.e., on the input data $\mathbf{x}$; measurement of these correlations (discussed in more detail around Eq.~\ref{eq:cov_matrix_elements} below) provides another source of nonlinearity in the reservoir feature map $f(\mathbf{x})$.\footnote{The nonlinearity from measuring correlations---or more precisely, the covariance matrix---intuitively comes about because correlations are defined by \emph{multiplications}, and a multiplication nonlinearly transforms the information in its operands so long as each operand contains information, i.e., is not just a constant.} 

The up-frequency converted optical state is subsequently detected by an EMCCD camera positioned after a diffraction grating. This configuration allows us to measure the photons in the output's frequency modes across $512$ pixels of the camera, measuring the different wavelengths that make up the converted state. The optical reservoir output can be measured by taking multiple frames with the camera where each frame is synchronized with each pulse of the AFC pump by triggering the camera to it.

We evaluated the classification accuracy of the resulting reservoir computer for different choices of the weak optical input (red box in Fig.~\ref{fig1}): squeezed-vacuum light, coherent light, thermal light, and supercontinuum light.

The squeezed light was a broadband frequency comb squeezed state spanning at least 1200\,nm through 1700\,nm \cite{presutti2024highly}. This state was generated by an optical parametric amplifier (OPA) based on a periodically poled lithium niobate (LN) waveguide. The waveguide was pumped by a $775$\,nm laser with a pulse duration of $200$\,fs, producing a vacuum-squeezed state spanning more than $400$ supermodes. The generated pulse is bright, with an average of approximately $700$ photons per pulse.

For each sample $\mathbf{x}$ to classify, from the recorded camera frames, we computed the photon-number covariance matrix $\Sigma$, whose elements are given by:

\begin{equation}
\label{eq:cov_matrix_elements}
 \Sigma_{ij} = \langle \hat{n}_i \hat{n}_j \rangle - \langle \hat{n}_i \rangle \langle \hat{n}_j \rangle
\end{equation}

\noindent where $\hat{n}=\hat{a}^\dagger\hat{a}$ is the photon number operator, with $i,j=1\dots M$ are the $M$ different frequency modes. Each camera frame contains the photon counts $n_i$ for each mode $i$, allowing construction of the covariance matrix $\Sigma$ from the set of recorded frames. Averaging the photon counts across the camera frames allows us to construct the mean-field vector $\bm{\mu}$ whose elements are $\bm{\mu}_i = \langle \hat{n}_i \rangle$. In all our experiments (unless otherwise noted), we performed the averaging to obtain $\bm{\mu}$ and $\Sigma$ using 5000 camera frames. This value was chosen as a balance between ensuring that the features $\bm{\mu}$ and $\Sigma$ are not too noisy, and enabling the experiments to be completed in a practical amount of time. 

We compare the classification performance by feeding the digital linear classifier with either the full covariance matrix $\Sigma$ (flattened to make it a vector) or the mean-field vector $\bm{\mu}$. The mean-field vector does not contain any information about the correlations between the measured modes. The covariance matrix, on the other hand, contains information both about the correlations (in its off-diagonal elements) and, implicitly, about the mean fields---because the diagonal elements describe the photon-number variance for each mode, and for states whose photon counts follow a Poisson distribution, the variance is the same as the mean.

\begin{figure*}[htb!]
	\includegraphics[width=0.99\textwidth]{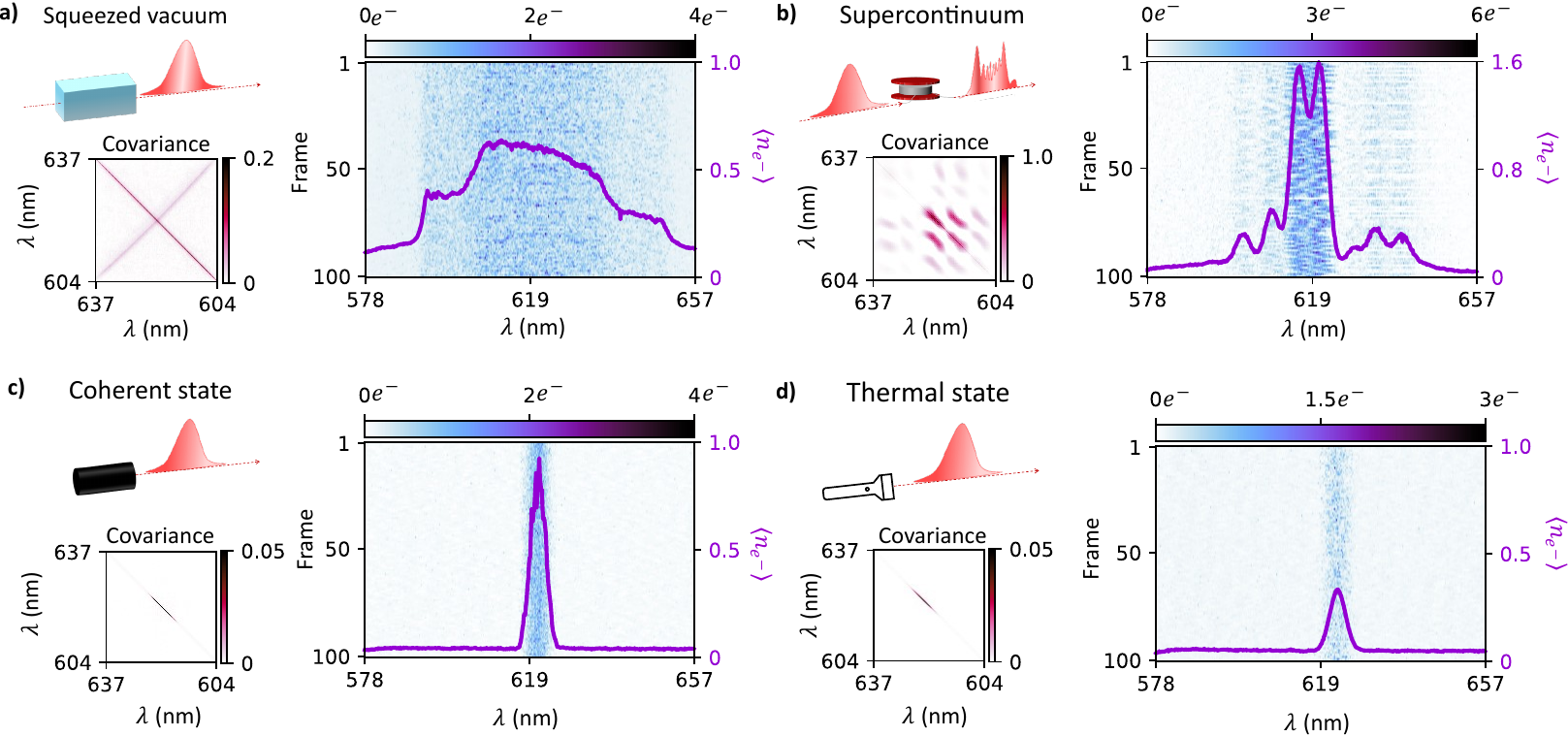}
	\caption{\textbf{Characterization of the behavior of the optical reservoir with quantum (squeezed) and classical (coherent, thermal, and supercontinuum) inputs, and with the AFC pump not encoding an input vector $\mathbf{x}$.} Comparison of measurement outcomes for different optically converted states. The experimentally measured camera frames are shown when the AFC pump consists solely of the chirped pulse without additional phase modulation, and the AFC signal input is from one of four different optical sources. Panels a)-d) display 100 consecutive frames across the 512 camera pixels for squeezed-vacuum (a) broadband supercontinuum (b), coherent (c), and thermal (d) states, respectively. The colorbar indicates the number of detected photons $n$ per pixel, while the purple line represents the mean field, computed as the average photon number per pixel across all measured frames. White dashed lines mark an arbitrary central pixel region used to calculate the covariance matrices among the overall detected frames, which are presented in panels e)-h) for the squeezed (e), broadband supercontinuum (f), coherent (g), and thermal (h) states, respectively.}
	\label{fig2}
\end{figure*}

Examples of the mean fields and covariance matrices that we observed are shown in Fig.~\ref{fig2} for each of the four types of input state we used; in all cases the AFC was pump did not encode an input vector $\mathbf{x}$ (or, equivalently, encoded a null vector).

\vspace{-0.1cm}

For the coherent light comparison, we employed a continuous-wave (CW) laser operating at a wavelength of $1550\,\text{nm}$ directly sent into the AFC crystal.
For the thermal light, we used the amplified spontaneous emission from an erbium-doped fiber amplifier (EDFA) also centered at $1550\,\text{nm}$, showing thermal light's statistical properties as demonstrated in Ref.~\cite{rosskopf2020ghost}.
We also wanted a comparison with broadband classical light (since our squeezed light is broadband, but neither the coherent nor thermal light sources were as broadband); for this we generated supercontinuum light in the near-infrared region by propagating a $1550\,\text{nm}$ pulsed laser through a highly nonlinear fiber (HNLF). Due to practical limitations in the equipment available to us, we were unable to have the classical-light sources (Fig.~1b--d) have identical spectra (and hence bandwidths) to the squeezed-light source (Fig.~1a); as we discuss in Sec.~\ref{sec:discussion}, a natural future step is to repeat the comparisons we report in this paper with light sources that are even more similar.

\section{Results}
\subsection{Application to tasks where the classes are not linearly separable}

\begin{figure*}[htb!] \includegraphics[width=\textwidth]{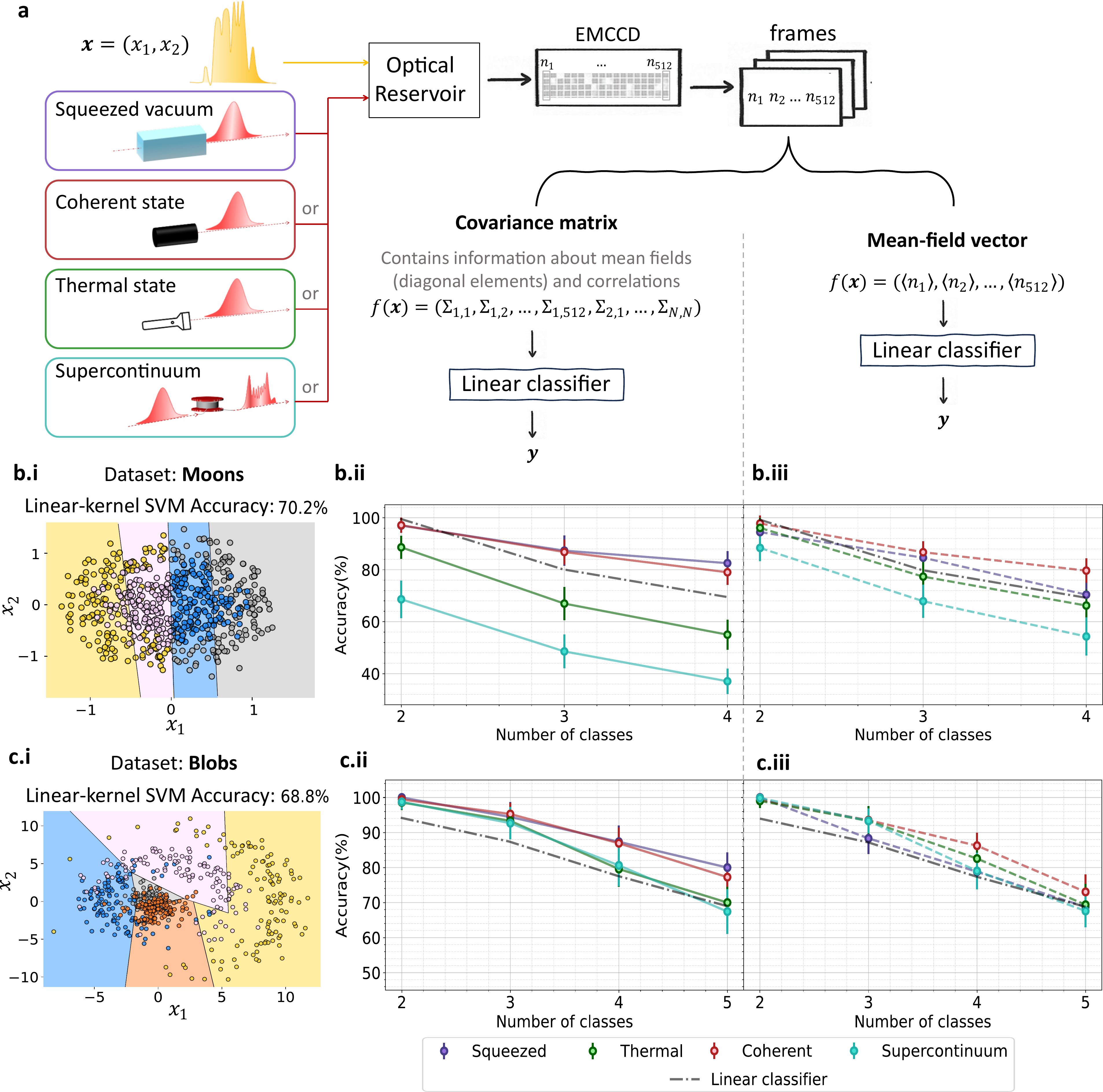}
\caption{\textbf{Experimental optical reservoir performance for non-linearly separable multi-class datasets.} Panel a shows a schematic of the implemented reservoir protocol. The phase of the AFC pump encodes the input features of the dataset under investigation. For each tested optical input, multiple EMCCD frames are collected, from which the covariance matrix and mean-field vectors are reconstructed. These serve as the two possible feature functions provided to the linear classifier. Panels b.i and c.i display the raw two features for the \emph{moons} and \emph{blobs} datasets, respectively. In the plots, each class is represented by a different color. The decision boundaries shown are obtained by applying the SVM with a linear kernel directly to the datasets, and the accuracy obtained on the test set is reported in the plot titles corresponding to the four-class and five-class classification tasks, respectively. This accuracy serves as a benchmark for linear classifier performance, represented by the dashed black line in panels b.ii, b.iii, c.ii, and c.iii, depending on the number of classes to distinguish. The test accuracy obtained between the linear classifier output and the dataset labels, as a function of the number of classes to distinguish for all the optical input states: squeezed-vacuum state (violet), coherent state (red), thermal state (green), and supercontinuum state (cyan), is reported in panels b.ii and c.ii when the linear classifier is trained with the experimental covariance matrix, or the mean field vector respectively. The error bars in the plots reflect the variability in performance when the dataset is shuffled, changing the composition of training and test sets while maintaining a fixed ratio.}
	\label{fig3}
\end{figure*}

\clearpage

We begin by evaluating the potential of the implemented optical reservoir to solve non-linearly separable multi-class tasks. Specifically, we assess the reservoir classification performance on two datasets generated using the \emph{moons} and \emph{blobs} functions in the Python \texttt{scikit-learn} package \cite{buitinck2013api}. These datasets are widely used in machine learning to benchmark classification algorithms due to their being simple examples of tasks that require nonlinearity \cite{guan2022novel,anai2024continuous}. 
We generated using these functions two-dimensional feature points associated with distinct labels. We randomly generated a total of 600 points associated with 4 and 5 different classes, as shown respectively in panels \textbf{a} and \textbf{b} of Fig.~\ref{fig3} where different colors refer to different classes. The dataset was then split into training and test sets, with $90\%$ of the points randomly selected for training and the remaining $10\%$ used for testing.

We started by employing as digital classifier a support vector machine (SVM) with a linear kernel. SVMs are a supervised learning algorithm that finds the optimal hyperplanes to separate the classes of the dataset, maximizing the distance between the hyperplane and the nearest data points from each class. For multi-class classification, we adopted the \emph{one-vs-one} approach, where separate binary classifiers are constructed for each pair of classes allowing the SVM to identify an optimal separating hyperplane between two classes at a time. For a given dataset with $n$ total samples $\{(\mathbf{x}^{(k)},l^{(k)})\}_{k=1}^n$, where $\mathbf{x}^{(k)}$ are the feature vectors and $l^{(k)}\in\{1,2,...,L\}$ are the associated labels, the algorithm builds $L(L-1)/2$ binary classifiers each trained to distinguish between two specific classes among the $L$ different classes.

Since we employed a linear classifier the kernel function simplifies to the dot product of the feature vectors $K(x^{(i)},x^{(j)})=\mathbf{x}^{{(i)}^T}\mathbf{x}^{(j)}$ thus allowing the SVM to learn a linear decision boundary in the feature space (Appendix \ref{App_opt}).

When applying the SVM classifier directly to the raw dataset, the algorithm cannot accurately separate the classes, as evident from the decision boundaries and the classification performance reported in the titles of panels \textbf{b.i} and \textbf{c.i}, and the dashed black lines in panels \textbf{b.ii}, \textbf{b.iii}, \textbf{c.ii} and \textbf{c.iii} of Fig.~\ref{fig3}.

When using the optical reservoir to process the raw features $\mathbf{x}$ before applying a linear classifier, the aim is to be able to achieve accuracies beyond those that a linear classifier can achieve on the raw feature vectors.

We quantified performance in terms of classification accuracy on the test set, i.e., the percentage of correctly classified samples. We studied the impact of different optical inputs (red input in Fig.~\ref{fig3}a) on classification accuracy, keeping the training procedure and linear classifier the same across the different choices to ensure a fair comparison.
We compare the performance for the different optical inputs each under two different conditions, as described in Sec.~\ref{sec:protocol}: one where the classifier is given the mean-field vector $\bm{\mu}$; the other where the classifier is trained on the covariance matrix $\Sigma$. Both $\bm{\mu}$ and $\Sigma$ were computed from $5,000$ recorded camera frames. For the mean-field-vector case, the training set consists of vectors with $512$ features for each dataset sample, corresponding to the number of camera pixels. For the covariance-matrix case, since the covariance matrix contains $512 \times 512 = 262,144$ elements, we apply feature selection to prevent overfitting due to the limited size of the dataset ($600$ total examples). Specifically, we retain only the most informative covariance elements, determined by their variance across the different samples. The number of selected features is optimized based on the highest accuracy achieved on a validation set comprising 8\% of the total dataset (see Appendix \ref{app_D} for details). This procedure typically resulted in selecting of the order of 10,000 features (elements of the covariance matrix) for the squeezed-light input, which correspond to the highly informative diagonal and anti-diagonal regions of the covariance matrix. For the supercontinuum light input, the optimal number increased to about 30,000 features. Conversely, for the coherent and thermal light inputs, where there were essentially no significant correlations between modes (off-diagonal matrix elements), only the meaningful diagonal elements of the covariance matrix were retained. The classification accuracies obtained for the moons and blobs datasets using the covariance matrix and the mean-field vectors are shown in Fig.~\ref{fig3} in Panels \textbf{b} and \textbf{c}. To ensure a fair comparison among the different optical inputs employed, we adjusted the optical power and the number of collected frames so that for each of the inputs, the average total number of photon counts were the same (see Appendix \ref{Appendix_resources}). The performance results are shown for the squeezed-vacuum (violet), coherent (red), thermal (green), and supercontinuum (cyan) inputs.

For both tasks, the highest accuracy was obtained when the optical input to the reservoir was the squeezed light and the covariance matrix was used as the reservoir output. For both tasks when considering the maximum number of classes, with the squeezed-light input there was an accuracy advantage of approximately 12 percentage points to using the covariance matrix instead of the mean-field vector. This indicates that useful information for distinguishing classes was encoded in the correlations between the measured modes.

For all the investigated classical optical inputs (i.e., coherent, thermal, and supercontinuum light), the use of the covariance matrix does not lead to any improvement over the use of the mean-field vector. The performance achieved with the covariance matrix was often worse than that achieved with the mean-field vector. An explanation for how this can occur is as follows. The diagonal of the covariance matrix contains the same information as the mean-field vector for light with Poisson statistics, so if the linear classifier learns to disregard all the information about the correlations in the covariance matrix (for the cases of classical optical inputs), then one would expect the reservoir computer achieve the same accuracies with the covariance matrix as with the mean-field vector. However, the training is done with a limited number of samples, and with a limited number of experimental shots, so the linear classifier may not learn to completely disregard the off-diagonal elements of the covariance matrix.   
Interestingly, while the supercontinuum light exhibited stronger (classical) correlations than we measured with the squeezed light (Fig.~2\textbf{a},\textbf{b}), these correlations apparently did not contribute any additional useful information for solving the classification tasks under study---as evidenced by the fact that in the case of supercontinuum light, the reservoir computer performed better when given only the mean-field vector rather than the covariance matrix.

\subsection{Application to practical machine-learning-benchmark tasks}
\label{Generic_subsec}

Having demonstrated the ability of the optical reservoir computer to outperform a purely linear classifier on tasks requiring nonlinearity, we experimentally evaluated its ability to perform two common machine-learning-benchmark tasks. Specifically, we evaluated the reservoir computer's performance on the spoken-vowels classification task, and on the MNIST handwritten-digits classification task.

The dataset for the vowels-classification task \cite{hillenbrand1995acoustic} contains 7 distinct classes, each represented by a 12-dimensional feature vector, related to the resonant frequencies in the vocal tract. The dataset includes $37$ examples per class, for a total of $259$ samples (details are provided in the Methods). The dataset features were encoded in the AFC pump beam using the same generic method previously described, without any special preprocessing.

\begin{figure*}[htb!]
    \includegraphics[width=0.8\textwidth]{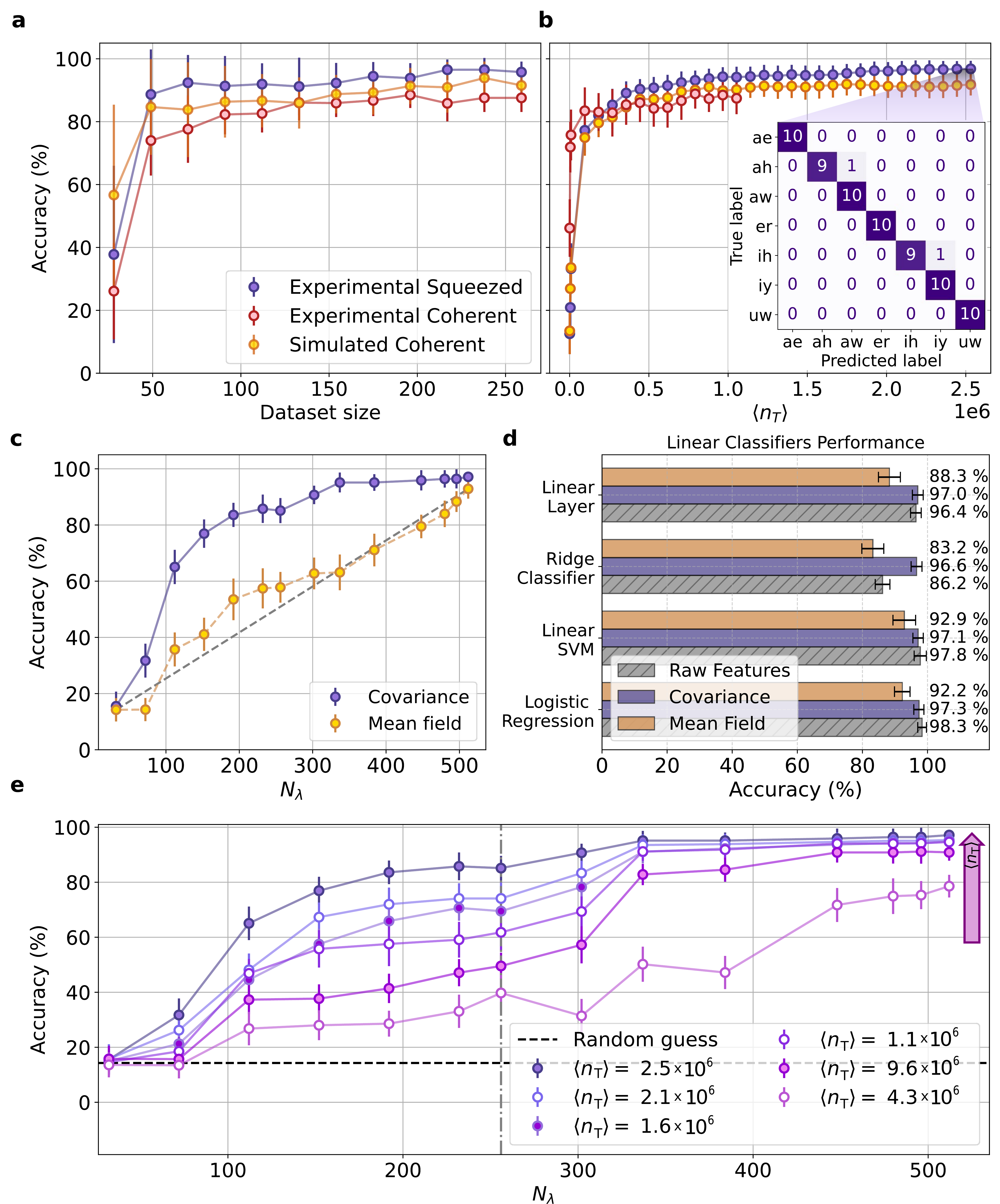}
	\caption{\textbf{Experimental optical reservoir performance on the spoken-vowel-classification task.} Panel \textbf{a} displays the classification accuracy on the test set as a function of the dataset size i.e. the total number of states, including both the training set (80\%  of the dataset) and the test set (20\%  of the dataset). for the squeezed-vacuum state (violet), coherent state (red), and simulated coherent state with the same mean field as the quantum state (orange). Panel \textbf{b} shows the performance as a function of the number of collected camera frames, which corresponds to an increase in the average number of collected photons $\langle n_T\rangle$, for both the squeezed-vacuum state (violet), experimental coherent state (red), and the simulated coherent state (orange). Results in both panels \textbf{a} and \textbf{b} are based on the use of the covariance matrix for all the optical states under investigation and each point reports training performed on measurement with the same number of photon counts. The inset of panel \textbf{b} presents the confusion matrix for the quantum state, illustrating the classification accuracy across the different classes. In panel \textbf{c} is reported the accuracy scaling with the number of modes $N_\lambda$ for the quantum reservoir (violet) and the classical reservoir (orange). Panel \textbf{d} shows a bar plot comparing the test accuracy of different linear classifiers applied directly to the raw dataset (i.e., without sending the data through the optical reservoir---giving the performance achieved classically by linear classifiers) (grey), the covariance-matrix reservoir output (violet), and the mean-field reservoir output (orange). Panel \textbf{e} classification accuracy as a function of the number of optical modes $N_\lambda$. Different colors in the plot correspond to data obtained with varying optical pump powers of the OPA, indicating the effect of reduced photon counts. The black dashed line is the accuracy of a random guess. The error bars in all panels represent the variability in performance resulting from shuffling the dataset, altering the composition of the training and test sets while keeping their ratio constant. The vertical gray line represents the point where exactly half the camera pixels were discarded.}
	\label{fig4}
\end{figure*}

We investigated the test accuracy as a function of the dataset size while keeping the ratio between training and test sets fixed. Panel \textbf{a} of Fig.~\ref{fig4} shows the performance achieved when training the model using the reconstructed covariance matrix for the multimode-squeezed state (violet) and for the coherent state (red). The plot also includes an orange curve representing the simulated performance of a broadband coherent state with the same average photon counts per pixel as the squeezed state. As with the synthetic tasks, here using squeezed light instead of coherent light resulted in the reservoir computer consistently achieving higher accuracies.

Increasing the number of photons in the optical state allows for a less-noisy estimate of correlations between the different frequency modes, and a growing gap in classification accuracy between squeezed and coherent light, as shown in Panel~\textbf{b} of the same figure. Each point in the plot corresponds to measurements made with states having the same average photon count per sample. The overall photon counts are increased by collecting more frames with the EMCCD camera.
When using the overall dataset and $30,0000$ frames, the quantum reservoir computer achieves a classification accuracy of $97\%$ on the vowel dataset, as shown in the confusion matrix in the inset.

We verified that the performance is robust to the specific choice of linear classifier at the output of the reservoir: in addition to evaluating a linear SVM, we tested a single neural-network layer with a linear activation function (trained using the Adam optimizer), as well as a Ridge classifier, and a logistic regression model. We evaluated the performance of these classifiers when applied directly to the raw dataset (i.e., without passing the data through the reservoir), the measured mean fields, and the measured covariance matrix. The results, shown in Panel~\textbf{d} of Fig.~\ref{fig4}, indicate that when training the models using the covariance matrix, the classification accuracy remains robust across the choice of the classifier. The violet bar in the plot represents the achieved test accuracy using the experimental covariance matrices for different classifiers, showing the same results within the error bars.
In contrast, when applying the classifiers directly to the raw dataset or to the mean-field measurements, significant variations in performance were observed. For instance, the Ridge classifier showed a considerable drop in accuracy compared to the SVM or neural network when trained on the raw or mean field data. Importantly, for each dataset and classifier, we optimized the model's hyperparameters individually, ensuring a fair comparison. The hyperparameters were not fixed but instead selected through cross-validation, allowing each model to operate at its best. This careful optimization process further emphasizes the robustness of the results: regardless of classifier type or optimization strategy, using the covariance matrix as the input to the classifier consistently yielded superior accuracy.

To study the influence of the number of modes used in the optical reservoir (for the case of squeezed inputs), we experimentally investigated the scaling of the reservoir computer's performance with the number of optical modes that were read out. To do this, we selectively removed information from the detected camera frames for a chosen subset of wavelengths, thus reducing the total number of informative pixels (see Appendix~\ref{Appedix_dimensionality}). This was done by replacing the photon counts in such regions with random values sampled from a Poisson distribution centered on the detected mean photon number. This procedure erases the information on the photon correlations related to the dataset from the selected window of pixels while preserving their statistical characteristics allowing us to emulate the behavior of a reservoir with fewer optical modes. Moreover, fixing the output vector dimensionality avoids introducing any undesired effects due to mere numerical issues in the training step. (The same study was done using an alternative procedure: fixing a number $N_\lambda$ of meaningful optical modes and substituting the rest with noise; the obtained results are very similar and are reported in Appendix~\ref{Appedix_dimensionality}.) Once this substitution was performed, we reconstructed the covariance matrix and proceeded with training the model using our standard procedure. In \textbf{c} of Fig.\ref{fig4}, we report the classification accuracy on the test set where we have progressively increased the number of optical modes $N_\lambda$.

\clearpage

We now compare the results obtained by scaling the number of modes, when using mean-field vectors versus using covariance matrices as inputs to the linear classifier. In the case of the mean-field vectors, we observe an approximately linear scaling with the number of modes. On the other hand, when the classifier is given access to the covariance matrix, the accuracy initially increases superlinearly with the number of modes. We can ascribe this as a benefit from information present in the correlations between modes. In part because the vowels-classification task is simple (a linear classifier can achieve 98\% accuracy on it), the superlinear scaling of accuracy with number of modes turns into a sublinear scaling as the accuracy converges to a plateau of ${\sim}97\%$ at $N_\lambda < 350$. With the same number of modes, using only mean-field vectors yields an accuracy more than 15 percentage points lower. 

We also investigated the impact of reducing the overall number of photons in the squeezed state. This was achieved by lowering the pump power of the OPA, which simultaneously reduces the number of squeezed photons and the strength of the measurable correlations. The results, as shown in panel e of Fig.~\ref{fig4}, reveal a clear dependence of the QRC's accuracy on both the number of informative optical modes used for the estimation of the covariance matrix and the total photon count in the state: in both cases, more was better.

Finally, we evaluated the performance of the optical reservoir on the MNIST digit-recognition dataset \cite{deng2012mnist}, a standard benchmark in machine learning.
The dataset comprises 28$\times$28 pixel grayscale images of handwritten digits (0--9), forming a classification task across ten distinct classes (Fig.~5\textbf{a}). To efficiently encode the two-dimensional images into the optical reservoir, we applied Principal Component Analysis (PCA) for dimensionality reduction, retaining the 100 most informative components (Fig.~5\textbf{b}). This reduced each sample to a 100-dimensional vector $\mathbf{x}$, which was input to the reservoir through modulation of the pump beam as with all the previous experiments.
We used a subset of the dataset, consisting of 300 samples for each of the 10 digits. We limited data collection to a total of $5000$ EMCCD frames; this corresponded to an average total photon count, among the different samples, of $3.6\times 10^5$, with the aim to balance data acquisition time with photon budget. The experiments we conducted with the MNIST dataset show the ability of our optical reservoir to handle high-dimensional datasets that, to the best of our knowledge, have dimensions well beyond what can be addressed by the quantum-optical reservoir computers that have been reported in the literature to date.

\begin{figure*}[ht!]
    \includegraphics[width=\textwidth]
    {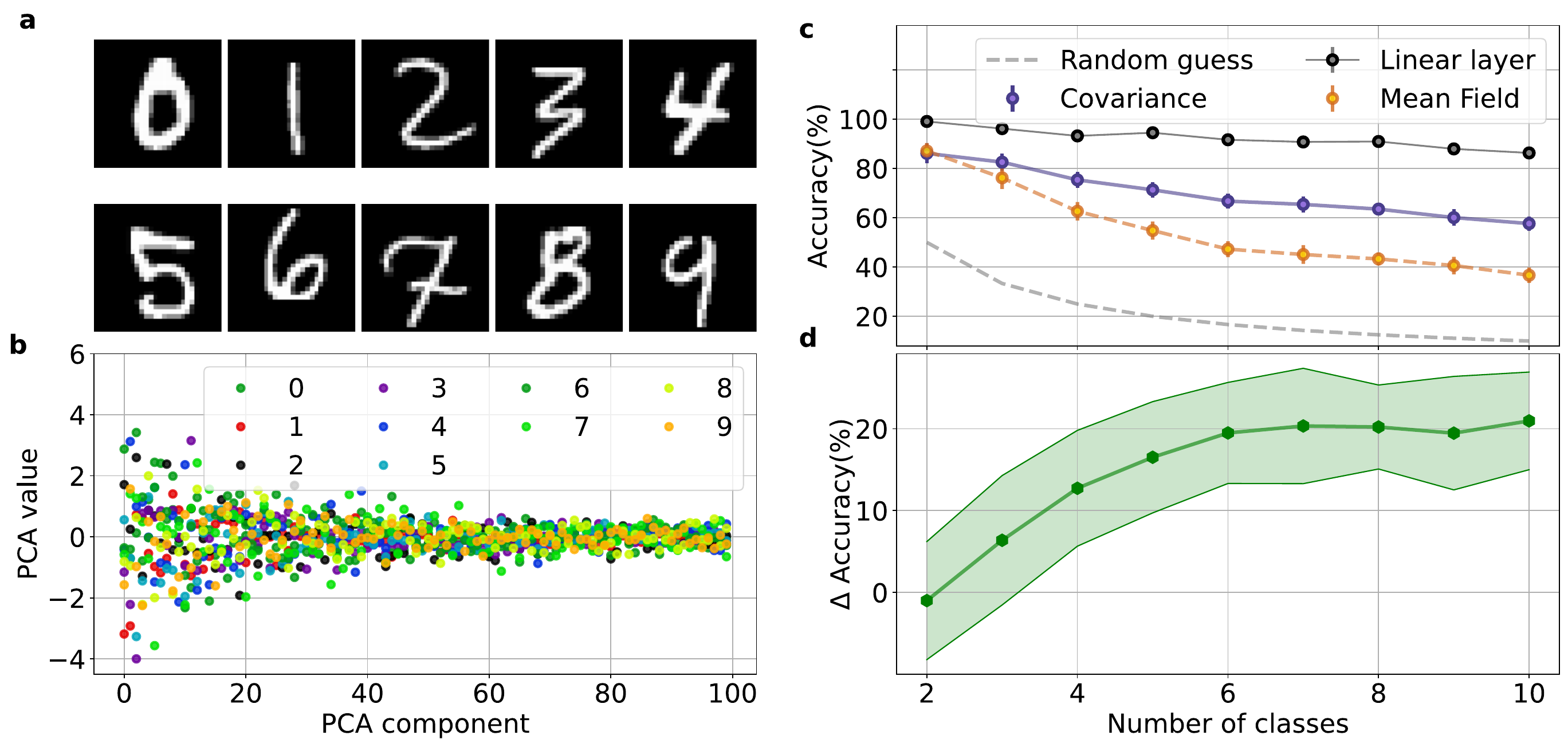}
	\caption{\textbf{Experimental optical reservoir performance on a task with high-dimensional features: MNIST handwritten-digit classification.} \textbf{a} Sample images from the 10 classes in the MNIST dataset \cite{deng2012mnist}. \textbf{b} The first 100 principal component values corresponding to the digits shown in panel \textbf{a}. The PCA values were used as the inputs $\mathbf{x}$ to the reservoir. \textbf{c} Test set accuracy as a function of the number of classes, comparing results from training using the covariance matrix (solid violet line) and mean-field measurements (dashed orange line). \textbf{d} The difference in accuracy between when the output layer has access to the covariance matrix (containing information about both the correlations and the mean fields) and when it merely has access to the mean fields, plotted as a function of the number of classes to discriminate. The shaded area represents the one-standard-deviation interval.}
	\label{fig5}
\end{figure*}

The results of our MNIST experiments are shown in Fig.~\ref{fig5}\textbf{c},\textbf{d}. We assessed the classification performance by gradually increasing the number of digits to be identified, starting with binary classification (digits 0 and 1) and progressively including more digits (0, 1, 2, etc.). We evaluated the accuracy both when providing the digital linear classifier with the covariance matrix and with the mean-field vector, and also evaluated the accuracy when the original feature vectors $\mathbf{x}$ were fed directly to a linear classifier without being passed through the reservoir. The performance gap between the reservoir computer using only the mean-field vector versus using the covariance matrix widened with increasing task complexity, again showing the importance of correlations in the reservoir output.

The obtained accuracy on the 10-digit classification task is approximately $60\%$, consistent with previous QRC results on the same task reported in Ref.~\cite{kornjavca2024large}. This relatively modest accuracy can be attributed to the limited size of the training set (3000 samples) and the constrained photon budget due to the limited number of collected camera frames (i.e., runs of the data through the reservoir for a single sample). We chose the number of frames (runs) based on wanting to keep the total data-collection time for the experiment to less than 24 hours, since nontrivial drift in alignment of the experimental apparatus can occur on longer timescales and would necessitate retraining, which we wanted to avoid.

\section{Discussion and Outlook}
\label{sec:discussion}

\subsection{Summary of results}

In this work, we have experimentally demonstrated the use of an optical system with $>400$ frequency modes to perform reservoir computing on a variety of classical machine-learning-classification tasks. We performed the experiments by operating the optical system with quantum (squeezed) and---for a subset of tasks---with classical (coherent, thermal, and supercontinuum) light sources, allowing a comparison of the computational performance of the reservoir computer across quantum and classical modes of operation. We tested both synthetic tasks that were designed to require nonlinearity to achieve high accuracy, and canonical machine-learning benchmarking tasks (spoken-vowel classification and MNIST handwritten-digit classification). For the synthetic tasks, the reservoir computer based on the optical system using squeezed light achieved accuracy substantially beyond that of a single-layer linear classifier, and also achieved high accuracy ($>$97\%) on spoken-vowel classification. While it was able to substantially outperform random guessing on MNIST handwritten-digit classification, the reservoir computer's accuracy was substantially lower than that of a single-layer linear classifier. Across all tasks, we observed that allowing the output layer of the reservoir computer (using squeezed light) access to the photon-number correlations between the measured modes of the optical reservoir resulted in the same or higher accuracy than providing access to the mean photon numbers only, and in several cases resulted in an advantage of greater than 20 percentage points. This provides experimental evidence in support of theoretical predictions that access to correlations enhances the power of quantum reservoir computers \cite{fujii2017harnessing,Wrightpreprint,nokkala2021gaussian}.

\subsection{Comparison between the use of quantum and classical light sources, and the role of correlations}

We ran the synthetic tasks requiring nonlinearity using both the quantum and the classical light sources and found that the quantum reservoir consistently yielded the best accuracy (either tied first or outright first, depending on the complexity of the task---the number of classes), provided that the photon correlations were fed to the output layer. The use of the photon correlations in the quantum experiments was essential for the reservoir computer to outperform the experiments where the reservoir was operated with classical, coherent light. This provides an interesting clue about the necessary ingredients for a quantum reservoir to outperform a similar but classical reservoir and is consistent with the hypothesis that a quantum reservoir using multimode squeezed light should be able to capture more useful or sophisticated features about the input data in its correlations than a reservoir using classical light.

However, since our experiments with quantum and classical light did not use light that was identical in all ways except for whether there was squeezing or not (especially the bandwidth of the input light was different for each light source), we are cautious to not over-interpret our results: we have observed that correlations in the squeezed-light case allowed the quantum reservoir computer to outperform the all the cases of classical reservoir computer we evaluated using various classical light sources, but it is possible that other classical light sources might result in higher accuracy still. For the vowels-classification task, we also presented results from simulations of the reservoir computer assuming a coherent-light source that had the same bandwidth as our experimental squeezed-light source---and our experimental results with the squeezed-light source showed higher accuracies than even these simulation results with the same-bandwidth coherent-light source. This provides further evidence suggesting that using squeezed light may consistently give superior accuracies versus using classical light sources.

Furthermore, the gap in accuracy between the squeezed-light case and one of the classical-light cases (coherent light) was present but within error bars (the squeezed-light case always outperformed the thermal and supercontinuum classical light cases by a large gap, well beyond that of the error bars). It is an important direction for future work to establish whether the squeezed-light case in our experimental protocol and setup does give an advantage in accuracy over all practical classical light sources (and if so, for which tasks). Another important limitation in our comparison experiments is that the effective optical loss between the squeezed-light generation and the detection (including sub-unity detector quantum efficiency) was sufficiently high (see Methods~\ref{methods_a}) that while the squeezed light did exhibit correlations close to the quantum limit, the correlations were not strong enough to rule out the possibility that they could in principle be generated by a classical light source. This limitation means that it would be inappropriate to ascribe our results---where the operation of the reservoir with squeezed-light yielded better accuracies than with the other, classical, sources of light---as definitively arising from the quantum nature of the squeezed light. Because the correlations between the output modes of the reservoir were important in achieving high accuracy in our squeezed-light experiments, it would be interesting to study the role of correlations in an even-lower-loss experimental setup where the correlations have an unambiguously quantum nature.

\subsection{Scaling of accuracy with the number of modes and with optical power}

For the vowel-classification task, we evaluated the accuracy as a function of how many frequency modes we read out from the reservoir. We found that the accuracy increased as we increased the number of modes read out---showing that the quantum reservoir computer was using the high dimensionality of the multimode optical system. We also varied the power of the pump of the squeezed-light source (and as a result the mean total number of detected photons) and observed that higher total numbers of photons led to higher accuracies, as one might intuitively expect, and that regardless of the optical power used, it was advantageous to include more measured modes in the reservoir's output rather than fewer.

We focused on comparing the quantum and classical modes of operation, both for the synthetic tasks and for the vowel-classification task, in the setting where the input optical powers were chosen such that the mean total detected number of photons was the same across the different quantum and classical light sources. While this allowed for comparisons that highlight the benefit of squeezed light and correlations, from a practical perspective this benefit under the constraint of a certain total number of detected photons is not automatically useful since it is easy to increase the brightness of the classical light sources but it is difficult to substantially increase the brightness of the squeezed-light source (by the metric of photons produced per pump pulse, the source we used already has record brightness \cite{presutti2024highly}). For the vowel-classification task, in our study of the accuracy as a function of the mean total detected number of photons, we found that within the range we tested, the accuracies obtained with the classical (coherent) light plateaued at values lower than those achieved with the quantum (squeezed) light. This leaves open the possibility that even with a much lower possible brightness, the squeezed-light case could outperform the coherent-light case---but further work with careful study of the accuracies at higher total photon numbers (for the coherent-light case) would be needed to definitively determine this. It would also be interesting to investigate the dependence of accuracy on the total photon number for different tasks, beyond vowel classification.

\subsection{Robustness to the choice of output layer}

For the vowel-classification task, we also tested four different variants of output layer (a standard linear layer, a ridge classifier, a linear support vector machine, and logistic regression) and observed that in each case, the quantum reservoir computing using photon correlations gave accuracies better than or equal (within error bars) to the classifier trained directly on the raw features (i.e., without passing them through the reservoir). This highlights the robustness of our results to choices one might make for the output layer.

\subsection{Comparison with the state-of-the-art QRC result}

Notably, the classification accuracy achieved by our reservoir computer using quantum (squeezed) light on the MNIST handwritten-digit classification task with all 10 digits $(58\pm 3)\%$ is comparable to that ($\approx$61\%) reported recently \cite{kornjavca2024large} using a state-of-the-art commercial neutral-atom-array quantum computer developed by QuEra. Beyond the fact that the accuracies can be directly compared because the benchmark task was the same (classifying MNIST digits), further details of the studies were also comparable: both studies adopted the same downsampling procedure as a preprocessing step, albeit with different numbers of retained components (8 in their case and 100 in ours), and the respective dataset sizes are of the same order of magnitude.

\subsection{Further limitations: classical simulability and conditions for quantum advantage}

As we described at the beginning of our paper, it is a grand challenge to find a practical application of GBS for which there exists a quantum advantage over all possible classical methods, and our work only addresses a part of this challenge: we have shown experimentally that GBS can be used for quantum reservoir computing, and that there is an advantage in using a squeezed-light source over using several classical modes of operating the same reservoir. However, it is likely that our experimental system (even when fed with squeezed light), with its nontrivial optical loss, is efficiently simulable classically \cite{oh2024classical}, especially because we only measure mean fields and two-body correlations \cite{bravyi2021classical}. Our work also does not address the crucial question of whether (and on which tasks) a GBS-based QRC could outperform all possible classical methods, including conventional digital-electronic classical methods, given the same resources (time, energy, money, etc.). It remains an open challenge---both theoretical and experimental---to design a QRC based on GBS that is efficient to operate (e.g., does not require exponentially many runs to compute the reservoir outputs) \cite{Wrightpreprint}, simultaneously is not efficient to simulate classically, and that outperforms classical methods on practically relevant tasks.

\subsection{Concluding remarks: future directions in quantum tasks}

While we have focused on classical machine-learning tasks in this work, our optical reservoir can in principle take in quantum light as an input to classify (for example, by replacing the squeezed-light input with the quantum light to be classified, and relying on the pump light entirely for driving the frequency-conversion process rather than for encoding the classical input to be classified). There is an expectation that quantum-machine-learning methods will in general be better-suited to solving quantum tasks rather than classical tasks \cite{cerezo2022challenges}; a natural direction would be to study the ability of the kind of optical reservoir we present to classifying quantum states of light \cite{mujal2021opportunities}.

The use of squeezed light in the optical system also inspires the question of whether the overall system we presented could be used for quantum sensing: squeezed light \cite{lawrie2019quantum} can provide an advantage in quantum sensing, so perhaps---in a special case of the above idea to apply the system to classifying quantum light---one could interact squeezed light with an object to be sensed and then feed this squeezed light (now having some encoding of the object in its phase or amplitude, potentially as a function of wavelength) into the optical reservoir. In this case, the objective may be for the reservoir computer to classify the object that was interacted with the squeezed light using as few photons as possible. This is an example of quantum computational sensing \cite{khan2025quantum}---the combination of quantum sensing with quantum computation.

We have, in this Discussion and Outlook section, suggested several avenues for further study of GBS-based QRCs on classical tasks, but the speculative applications to classifying quantum states might well be the first where a GBS-based QRC could deliver a quantum advantage.

\vspace{8ex}
\textit{Note added:} Ref.~\cite{joly2025harnessing} appeared as an arXiv preprint the day before we submitted our manuscript to the arXiv. It reports experiments realizing a photonic feedforward reservoir computer (i.e., an extreme learning machine) using a photon-pair source and propagation through a multimode fiber that acts as a random beamsplitter unitary transformation. The machine-learning task they experimentally tested is binary classification of two MNIST handwritten digits (`0' and `1'), which corresponds to the `Number of classes = 2' data point in Fig.~\ref{fig5}c in our work.

\section*{Data and code availability}

Experimental data and code to replicate the figures in this paper are available at \url{https://doi.org/10.5281/zenodo.15288036}.

\section*{Author contributions}

V.C. designed and carried out the experiments and data analysis, and performed the numerical simulations. M.M.S. contributed to the design of the experiments. F.P. and L.G.W. designed the experimental setup. F.P. built the experimental setup with assistance from L.G.W., B.K.M., S.-Y.M., and T.W.. R.Y. and T.O. provided advice on quantum-optical modeling of the system. V.C. and P.L.M. wrote the manuscript with input from all authors. L.G.W. and P.L.M. conceived the project, and P.L.M. supervised the project.

\section*{Acknowledgements}

We thank NTT Research for their financial and technical support. Portions of this work were supported by the National Science Foundation (award CCF-1918549). V.C. acknowledges support from PNRR MUR Project No. PE0000023-NQSTI (Spoke 4), FARE Ricerca in Italia QU-DICE Grant n. R20TRHTSPA, and from the Fulbright Research Scholar Program. P.L.M acknowledges support from a David and Lucile Packard Foundation Fellowship and membership of the CIFAR Quantum Information Science Program as an Azrieli Global Scholar. B.K.M. was supported by the Intelligence Community Postdoctoral Research Fellowship Program at Cornell University, administered by Oak Ridge Institute for Science and Education through an interagency agreement between the U.S. Department of Energy and the Office of the Director of National Intelligence.

\section{Methods}
\label{sec:methods}

\subsection{Experimental setup}
\label{methods_a}

Squeezed light is generated by pumping a nonlinear crystal, seeded with vacuum, using ultrafast pulses with a duration of $200$ fs. The OPA produces multimode-squeezed-vacuum light with a high photon flux. The properties of this multimode-squeezed light are analyzed using the Bloch-Messiah decomposition, allowing for the quantification of squeezing and anti-squeezing levels associated with the different supermodes. As reported in Ref.\cite{presutti2024highly}
by fitting the average number of detected photons as a function of the OPA pump power, we estimate the presence of approximately $430$ supermodes. Assuming that all supermodes are equally squeezed, it is possible to infer a lower bound for the squeezing level of each mode, which is at least 3 dB. This estimate is based on an average detection efficiency of $71\%$.

If we consider the overall optical transmission of the system, it is approximately $60\%$, factoring in multiple sources of loss. These include the camera's quantum efficiency, which is $95\%$ at $620$ nm. Notably, when considering different dataset examples the phase modulation on the AFC pump introduces additional losses, which can vary by up to $20\%$ depending on the specific dataset example being considered.
Consequently, not all collected states exhibit squeezing below the shot-noise limit by the time they reach the camera. 
However, residual correlations between modes —originating from the initial quantum properties of the squeezed state— persist despite losses as evident from the reconstructed covariance matrices. These correlations, while no longer strictly quantum at the detection stage, retain a structured nature that is crucial for encoding relationships in the data. These intra-mode correlations play a critical role in the system's ability to achieve superior classification performance compared to different light sources. For more detailed studies on the experiment and the generation of squeezed light we refer to Ref.~\cite{presutti2024highly}.

\subsubsection{Pulse shaping and frequency conversion}
The crystal used for frequency conversion is a 3 cm poled KTP crystal (Raicol Crystals).
The poling is designed for broadband sum-frequency generation, where the central wavelengths are $1550 + 1033 \rightarrow 620$~nm, and the 1033~nm is the pump wavelength.
The spatial frequency of the poling profile varies linearly over the length of the crystal, in order to phase match a broader frequency bandwidth.
The crystal is designed for use at 48\textdegree C and held in an oven (Eksma Optics HP30).

The 1033~nm pump laser is an Amplitude Satsuma, which is temporally shaped prior to the AFC crystal.
The pulse shaper is designed as follows.
The beam is diffracted by a transmission grating (Ibsen Photonics PCG-1765-808-981) and focused by a 150~mm cylindrical lens (Thorlabs LJ1629L2-B) onto an spatial light modulator (SLM; Meadowlark P1920-0600-1300-PCIe).
A vertically-oriented blazed grating is written on the SLM (orthogonal to the frequency-dispersion due to the diffraction grating).
The spatial phases (vertical translations) of the blazed grating along the wavelength axis impart spectral phases to the pulses.
The 1st order reflection is recombined as it travels back through the cylindrical lens and grating and is directed towards the AFC crystal.

\subsubsection{Spectrometer and detection}

The visible light spectrometer is based on a diffraction grating (Ibsen PCG-1908/675-972) imaged by an objective lens (Olympus UPLFLN4x) onto the N\"uV\"u HN\"u 512 IS EMCCD camera.
The 620~nm light from the AFC is measured on the camera after appropriate filtering of the pump (Thorlabs DMLP900; Semrock FF01-632/148).
A nominal EM gain of 3000 is used on the camera, and the blanking and exposure times are set to 0.1~ms.
The camera is triggered by the Amplitude Satsuma laser, with an appropriate time delay (IDQuantique ID900), to ensure that each frame corresponds to a single pulse.
The laser repetition rate of the Satsuma is chosen to match the maximum frame-rate of the camera for the given region of interest.

The following summarizes the EMCCD camera photo-detection and signal acquisition process.
Each photon incident on and absorbed by the camera's CCD generates a photo-electron.
The electron multiplication (EM) gain amplifies this few-electron signal to a sufficiently large voltage that can be read out by the analog-to-digital converter, after a constant bias is applied.
By subtracting the bias and dividing by the total gain---the EM gain and analog-to-digital conversion factor---the camera digital readout (``pixel value'') can be approximately converted back into the number of photo-electrons, hence the number of captured photons.
The EM gain process is stochastic: this allows us to determine the average number of incident photons per pixel, but not the number of photons in any given single frame.

\subsubsection{Light sources}

The OPA waveguide is a 4~cm MgO-doped LN ridge waveguide (Covesion WG-1550-40 WGCK40), optimized for wavelength conversion from 1550 nm to 775 nm. It is engineered to support single spatial modes at near-infrared (near-IR) wavelengths. This device is mounted in a temperature-controlled oven (Covesion PV40) to ensure optimal performance. Additionally, the waveguide serves as a common spatial mode for all light sources, which propagate through it prior to the adiabatic frequency conversion (AFC) process.

Squeezed light is generated by vacuum-seeding the OPA with a pulsed 775 nm pump beam. The 775 nm light is produced by cascading second harmonic generation (SHG) and an OPA process in lithium triborate (LBO) crystals (Newlight Photonics). Both crystals are temperature-tuned and non-critically phase-matched to optimize the conversion efficiency. The crystal lengths are carefully chosen to match the temporal walk-off of the 200 fs pulses, preserving the pulse duration and efficient interaction.

The first crystal generates 517~nm from 1033~nm.
The second carries out the 517~nm~$-$~1548.5~nm~$\rightarrow$~775~nm process.
The 1548.5~nm seed is generated by a diode laser (ILX Lightwave 79800C), amplitude-modulated by a stabilized (Oz Optics MBC-SUPER-PD-3A) electro-optic modulator (EOM) (Eospace AZ-DS5-10-PFA-PFA-LV-LR), to generate 10~ns square pulses.
These seed pulses are optically amplified (Pritel PMFA-20), following an isolator.
The EOM pulsing is triggered by the Amplitude Satsuma 1033~nm laser, with an appropriate time delay (IDQuantique ID900).
The EOM is driven by a pulse/function generator (HP-8116A).

The supercontinuum is generated by pumping a highly-nonlinear fiber (Thorlabs HN1550) with mode-locked 1550~nm laser (NKT Origami).
These pulses were not temporally synchronized to the Amplitude Satsuma laser, hence the frames with little or no overlap between the AFC pump and supercontinuum were removed in post-processing.

The coherent laser light consisted of a 1550~nm continuous-wave laser (JDSU mTLG-C1C1L1) appropriately attenuated to match the desired flux upon conversion in the AFC.

An unseeded EDFA (JDSU MEDFA-A1-Z001) was used as a thermal light source, with appropriate attenuation, as amplified spontaneous emission is a thermal process.

\subsection{Phase encoding of dataset features}

The features of the dataset are encoded into the optical phase of the AFC pump pulse through the SLM. This phase encoding method is consistent across all the datasets we investigate, with the only variation being a constant scaling factor applied to the dataset features. This factor ensures that the value set on each SLM pixel stays below a defined threshold, preventing the generated pulse from having controlled intensity peaks. The phase modulation occurs over a single row of 1920 pixels.
Given the grating resolution and the fact that the pump pulse covers approximately $40\%$ of the pixels, we can effectively manipulate the phase of around $\sim 100$ modes.

\begin{figure*}[!t]
	\includegraphics[width=0.99\textwidth]{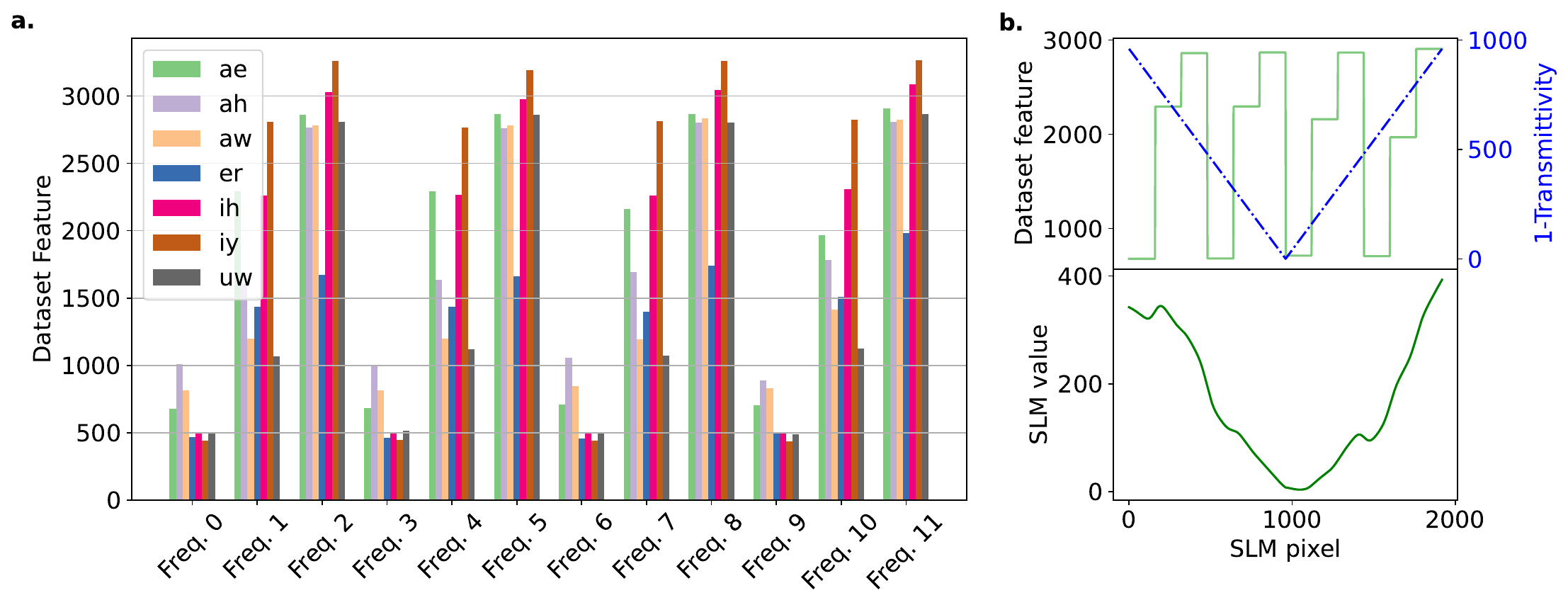}
	\caption{\textbf{Features encoding of the vowels dataset.} a. Representation of the 12 resonant frequencies of the vowels dataset. Each color corresponds to one of the 7 different vowel classes. \textbf{b} We report how the 12 resonant frequencies are distributed across the SLM pixels, with each feature replicated across multiple contiguous pixels to fully utilize the 1920 pixels available. The blue curve represents the compensating function applied to account for the non-uniform transmissivity of the SLM pixels, ensuring consistent phase modulation. The lower part of panel b. displays the resulting phase mask applied to the SLM pixels.}
	\label{figm_1}
\end{figure*}

To illustrate this encoding approach, we consider the vowel dataset as an example. The vowel dataset comprises $37$ examples for each of the $7$ classes, where each example is represented by $12$ resonant frequencies. An example from each class is shown in panel a of Fig.\ref{figm_1}. In order to fill in all the SLM pixels, we replicate each feature across several contiguous pixels, such that the entire dataset can be distributed across the 1920 pixels. Since the dataset contains 12 features per example, each feature is mapped to $160$ contiguous SLM pixels for this task, as depicted in panel b of Fig.~\ref{figm_1}, resulting always in a 1920 feature vector.

Since the SLM pixels transmissivity is not uniform, but central pixels have higher transmissivity compared to those at the edges, we introduced a compensating function. The function is reported in blue in the plot and it is multiplied to all the feature vectors to adjust for the difference in the pixels transmissivity, ensuring that the phase modulation is applied consistently across the pump pulse. To avoid huge changes between adjacent pixels we also implement a Gaussian filtering on the obtained vector. This filter averages adjacent values with weights determined by the Gaussian function with a selected standard deviation.
The obtained vector is then added to a fixed quadratic phase (chirp) to maintain the desired pump pulse duration. The final SLM values, for a randomly selected example from the vowel dataset, are reported in the lower part of panel b.

It is important to note that both the compensating function for the SLM transmissivity variation, the gaussian filtering, and the quadratic phase chirp remain identical for all the classification tasks we explore, keeping this generic encoding process independent both from the task and more importantly from the optical input employed.

\appendix

\section*{APPENDICES}

\section{Resources count}
\label{Appendix_resources}

\begin{figure*}[!htb]
	\includegraphics[width=0.99\textwidth]{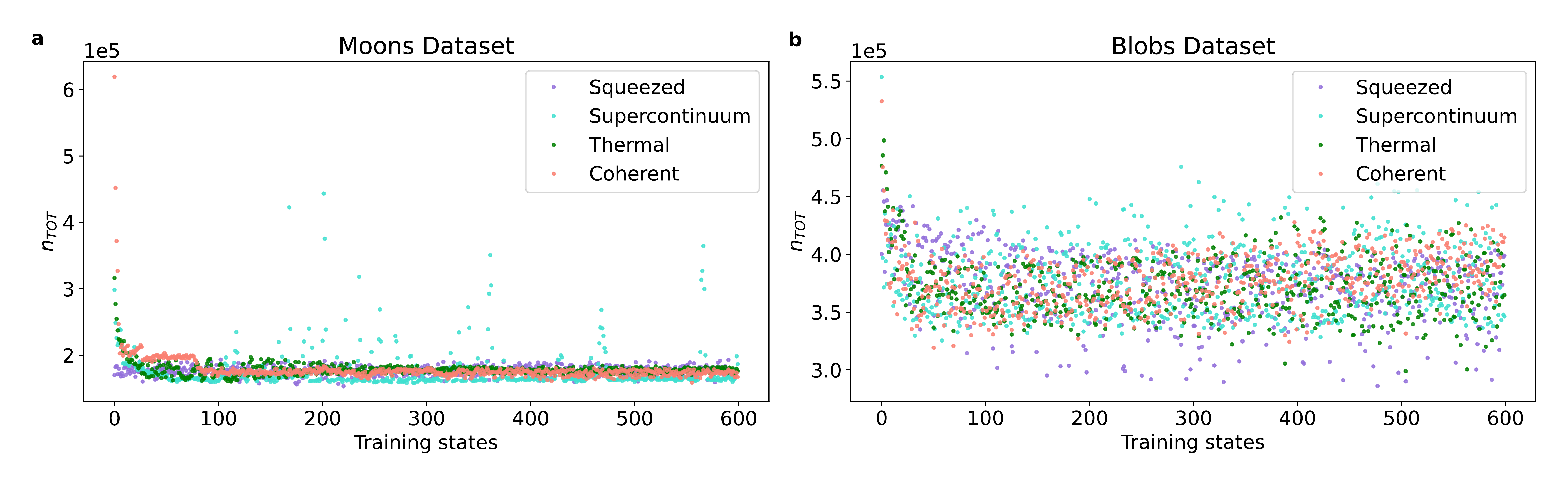}
	\caption{\textbf{Total photons counts per dataset example} Sum of the counts obtained among the 512 pixels and for the whole number of considered frames for all the $600$ samples of the moon dataset (\textbf{a.}) and the blobs dataset (\textbf{b.}). The total photon budged for each dataset example is reported for all the tested optical input: quantum light (blue), supercontinuum light (purple), thermal light (green) and coherent light (red).}
	\label{fig1a}
\end{figure*}

To ensure a fair comparison of the reservoir's performance across the different optical inputs in each of the studied scenarios, we carefully matched the total resource usage, quantified as the total number of detected photons. To this end, we adjusted the number of processed camera frames to equalize the overall photon budget. In Fig.\ref{fig1a} we report the detected photon counts for each of the dataset examples where we have selected the number of frames for each optical state to have the same total number of detected photons averaged among the $600$ total dataset examples. More specifically, the average number of counts among the $600$ examples of the moons dataset is $(1.76\pm 0.07)\cdot 10^5$ for quantum light, $(1.75\pm 0.30)\cdot 10^5$ for supercontinuum light, $(1.79\pm 0.10)\cdot 10^5$ for thermal light, and $(1.79\pm 0.25)\cdot 10^5$ for coherent light. While for the blobs dataset is $(3.7\pm 0.3)\cdot 10^5$ for quantum light, $(3.8\pm 0.3)\cdot 10^5$ for supercontinuum light, $(3.7\pm 0.3)\cdot 10^5$ for thermal light, and $(3.8\pm 0.2)\cdot 10^5$ for coherent light.

\section{Singular values analysis}

An alternative way to quantitatively assess which optical input contains the most information relevant for classifying the dataset is to perform a singular value decomposition (SVD) using principal component analysis (PCA) on the mean fields and covariance matrices of the measured data. Specifically, we first apply a standard scaling procedure to each dataset to ensure comparability, followed by PCA to extract the singular values. We carry out this procedure on the measurements retrieved for all four optical inputs: quantum light, coherent states, thermal states, and a broadband supercontinuum state, each using the same number of principal components. The singular values obtained from the PCA reflect the variance captured by each dataset, with larger singular values indicating a higher capacity to encode relevant information for classification.

\begin{figure*}[!htb]
	\includegraphics[width=0.99\textwidth]{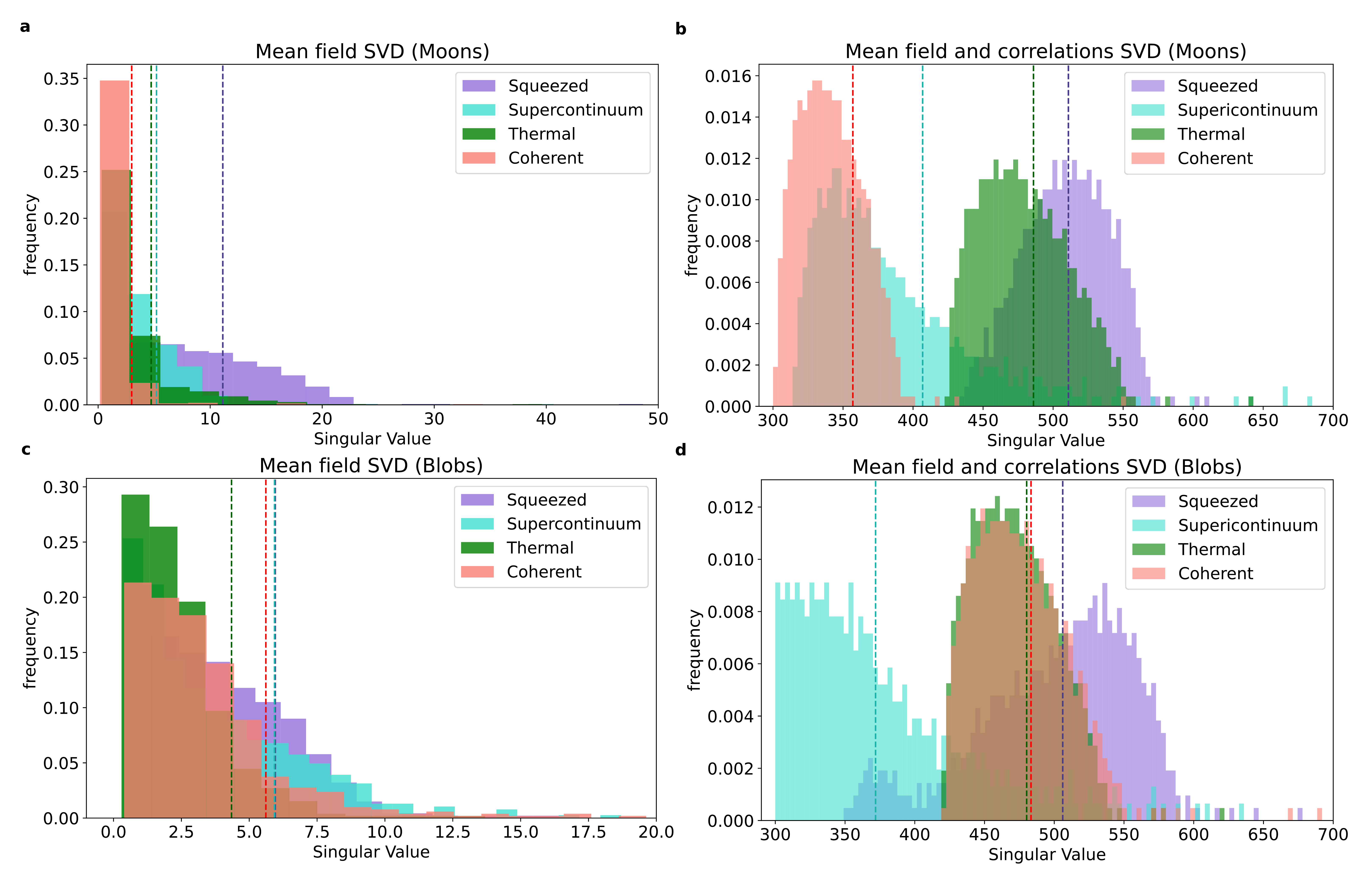}
	\caption{\textbf{Singular values decomposition analysis}. Singular values histograms obtained after performing the PCA analysis on the mean field vectors (a-c) and on the covariance matrices (b-d) for the moons and the blobs datasets. Each optical input's singular values are binned and plotted with normalized frequency. The dashed vertical lines represent the mean singular values for each optical input.}
	\label{fig2a}
\end{figure*}

The singular values obtained from the covariance matrices with the different optical states show that the measurements obtained after the conversion of quantum light have more significant features than the other optical inputs. This emerges by looking at the SVDs reported in Fig.\ref{fig2a} where the histogram of singular values for the four optical inputs is reported.
The quantum input shows a distinct distribution with higher mean singular values compared to the classical inputs, confirming the greater informational content derived from quantum correlations when using the covariance matrix. This relates to the ability of the quantum optical reservoir to outperform its classical counterparts, offering a clearer distinction between classes and thus improving classification accuracy.

This advantage does not hold anymore when considering only the mean field since the information about quantum correlations in this scenario is erased by averaging among the collected camera frames. When considering only the mean fields the SVDs extracted from the measurements with the different inputs are comparable and the observed differences are task dependent. For instance, we can see that for the blobs dataset, the singular values obtained from the mean field vector of all the optical converted states are comparable.

This approach is another valid way to compare the informational content of the optical reservoir with the different optical inputs, revealing the unique advantages provided by quantum correlations in the classification tasks.

\section{Reservoir dimensionality study}
\label{Appedix_dimensionality}

To analyze how the dimensionality of the quantum reservoir affects the test accuracy, we post-process the collected frames from the EMCCD camera by selectively reducing the number of informative pixels. This approach allows us to simulate the behavior of systems with fewer optical modes. Importantly, the information from the dataset is not distributed uniformly across the camera’s $512$ pixels. Therefore, we begin by examining the test accuracy of the model trained on covariance matrices reconstructed after introducing white Poissonian noise to a chosen subset of pixels. This subset can be varied in size and position across the 512-pixel array, enabling us to study the model's performance as a function of the initial pixel position where noise is introduced.

We expect that the pixels located near the center of the array, which correspond to regions with higher photon counts, will contain the most significant portion of the dataset’s information. Conversely, pixels at the edges, which generally register fewer photons, are expected to contribute less to the reservoir’s performance. As a result, the removal or substitution of these edge pixels with noise is likely to have a minimal effect on the overall performance. 
We report the accuracy as a function of the initial pixel location where noise is introduced, varying the size of the noise window. This enables us to identify the optical modes that encode the most information about the dataset.

\begin{figure*}[!htb]
	\includegraphics[width=0.99\textwidth]{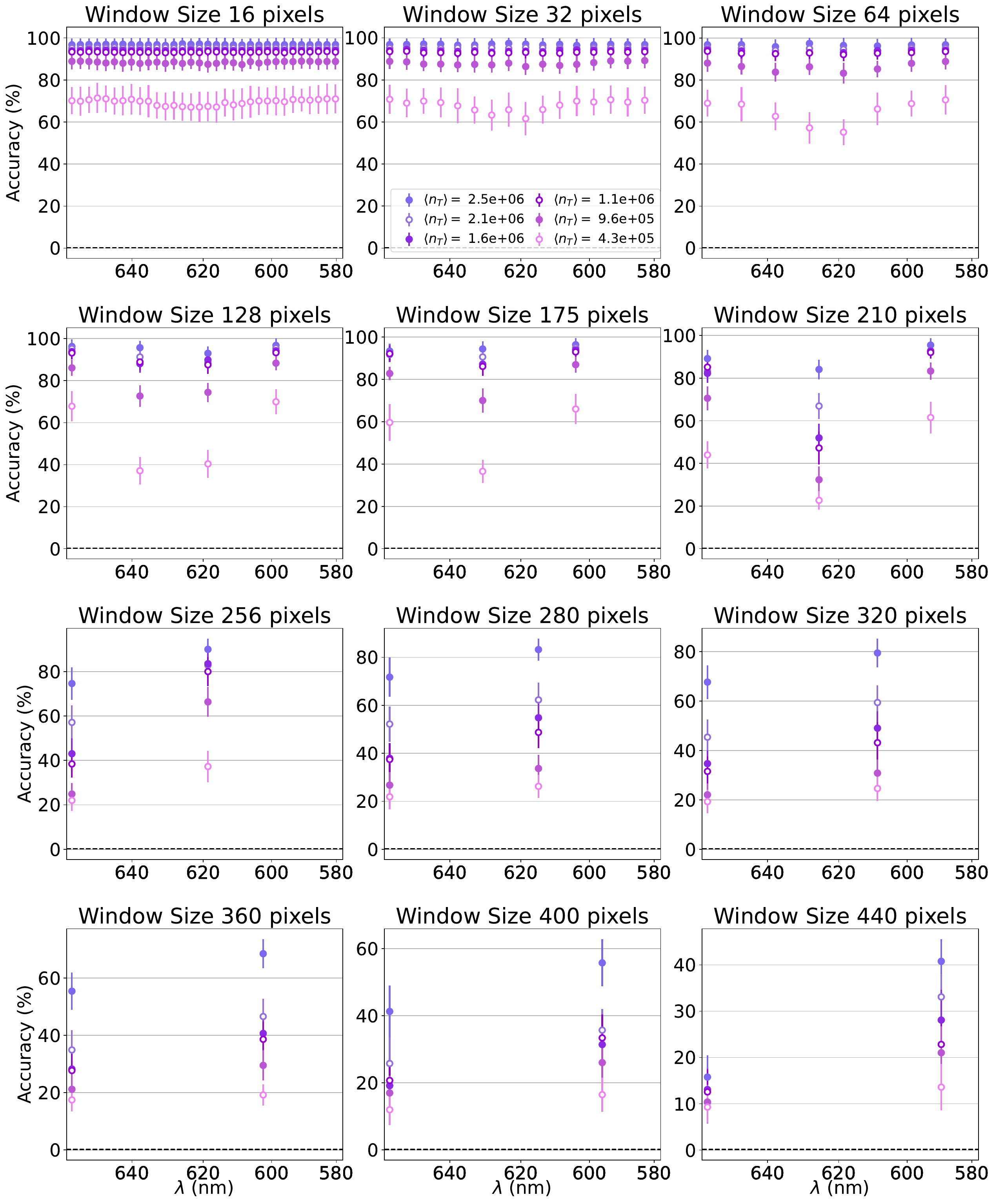}
	\caption{\textbf{Reservoir performance as a function of informative mode sizes.} Test accuracy for different noise window sizes changing the initial pixel substitute with Poissonian noise. The top set of plots refers to studies done with relatively small selected noise windows, from this it is evident that the central pixels region is the most informative since when those pixels are substituted with noise the accuracy drops. The different colors in the plots correspond to datasets collected at varying DOPA pump powers, resulting in quantum states with different overall photon counts. The horizontal dashed black line indicates the accuracy expected from random guessing, serving as a baseline for comparison.}
	\label{fig3a}
\end{figure*}

\clearpage

The removal of these highly informative pixels is crucial for accurately studying how the system's performance scales with dimensionality. Including non-informative regions in the analysis could indeed compromise the true trend. Once the most informative pixels are identified, we gradually increase the size of the noise window, targeting the regions with the highest information density. For each window size, we evaluate the model's classification performance as the number of noisy informative pixels increases, as shown in Fig.\ref{fig3a}.

\begin{figure*}[!htb]
	\includegraphics[width=0.99\textwidth]{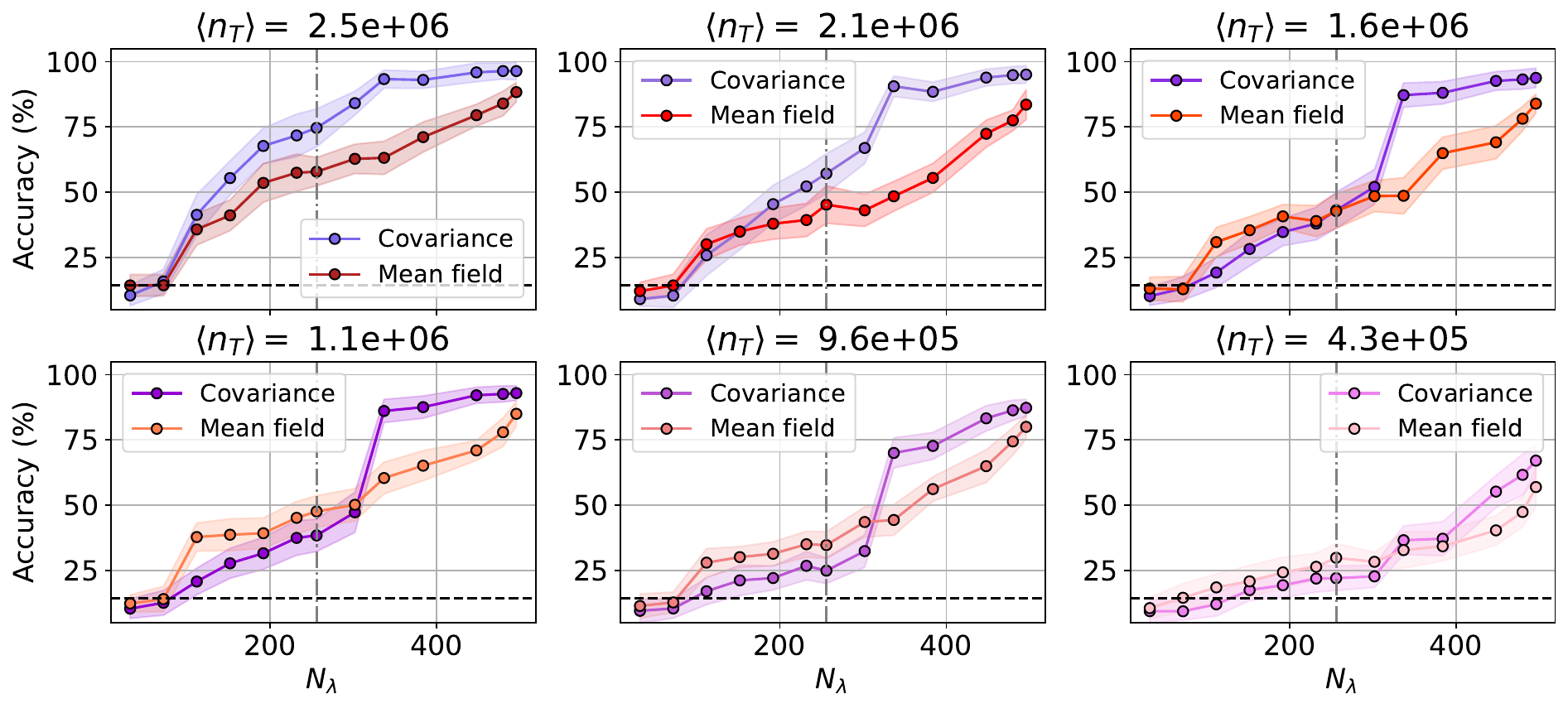}
	\caption{\textbf{Squeezed vs. coherent light reservoir dependence on system dimensionality}. Test accuracy comparison achieved using the covariance matrix (squeezed light reservoir, shown in violet) and the mean field (coherent light reservoir, shown in red), both computed from the same measurement datasets. The results are shown as a function of the increasing number of informative optical modes, which corresponds to increasing the number of central pixels considered in the analysis. The difference between the quantum and the classical reservoir becomes evident beyond the gray point-dashed vertical line, corresponding to windows with more than 256 informative pixels, where the measurements begin to reflect information about inter-mode correlations. The different subplots correspond to datasets collected at varying DOPA pump powers, resulting in quantum states with different overall photon counts. The horizontal dashed black line indicates the accuracy expected from random guessing, serving as a baseline for comparison.}
	\label{fig4a}
\end{figure*}

This approach allows us to explore the performance scaling with the number of optical modes, providing valuable insight into the dimensionality dependence of our quantum reservoir versus its classical counterpart. 
We study the performance scaling with the number of informative pixels when using the covariance matrix (capturing quantum correlations) and training with the mean field (mimicking a classical reservoir with the same average number of photons per mode), the results are reported in as shown in Fig.\ref{fig4a}. In the quantum case, the reservoir is composed of squeezed states, where photon pairs are emitted on two correlated frequency modes. Initially, the classical and quantum reservoirs exhibit similar behavior because more than half of the camera pixels are required to begin capturing the entanglement between modes. However, once a sufficient number of informative modes is included, allowing the quantum correlations to be detected, we start to observe a gap among the obtained performance. From this point, the quantum reservoir's performance begins to scale exponentially with the number of modes, leading to a rapid expansion of its computational capacity. This exponential growth is a result of the quantum reservoir's ability to encode the dataset features in the higher dimensional Hilbert space while the classical reservoir remains constrained to a linear scaling with the system dimensionality.
For this reason, the performance gap between the quantum and classical reservoirs starts to widen exponentially. However, this exponential scaling eventually plateaus, as the quantum system reaches the limit imposed by the inefficient dataset's feature encoding. Indeed, without tailoring the dataset encoding to the specific quantum state generated, it prevents further exponential growth. Nevertheless, as the number of modes increases, the quantum reservoir still retains a clear advantage in terms of performance than what can be obtained with a classical reservoir.

\begin{figure*}[!htb]
	\includegraphics[width=0.99\textwidth]{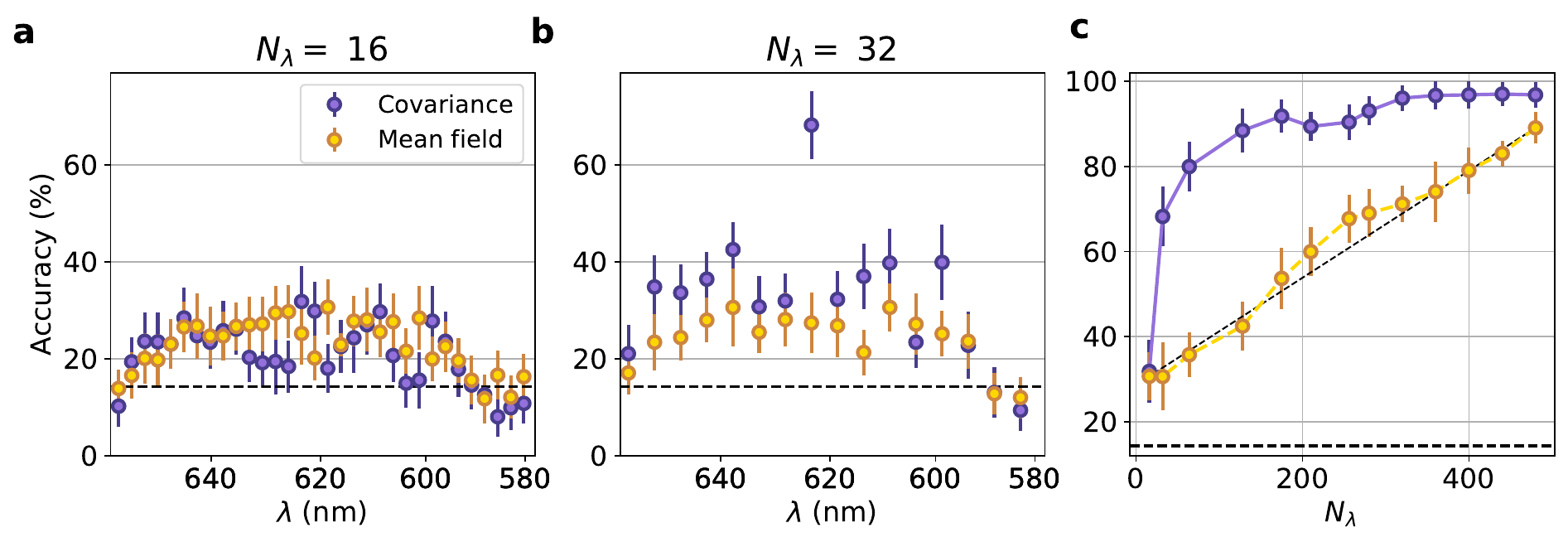}
	\caption{\textbf{Squeezed vs. coherent light reservoir dependence on system dimensionality}. Study of the reservoir performance for a subset of optical modes. Panel a and b show the test accuracy when selecting $N_\lambda$ modes starting from the wavelength reported on the x-axis. Panel c shows the performance for an increasing number of optical modes. In violet, the results achieved training and testing the linear classifier with the covariance matrices while in orange are reported the results obtained when using only the mean field vectors.}
	\label{fig5a}
\end{figure*}

The same study can also be carried out by selecting only a number of informative wavelengths $N_\lambda$. The results obtained when choosing different wavelength regions are reported in panels \textbf{a} and \textbf{b} of Fig.\ref{fig5a}. Also in this case we observe the different scaling (panel \textbf{c} of the same figure) with the system dimensionality when comparing the performance achieved with the quantum reservoir and the classical one.

\section{Measurements and feature selection}
\label{app_D}

\begin{figure*}[!htb]
	\includegraphics[width=0.99\textwidth]{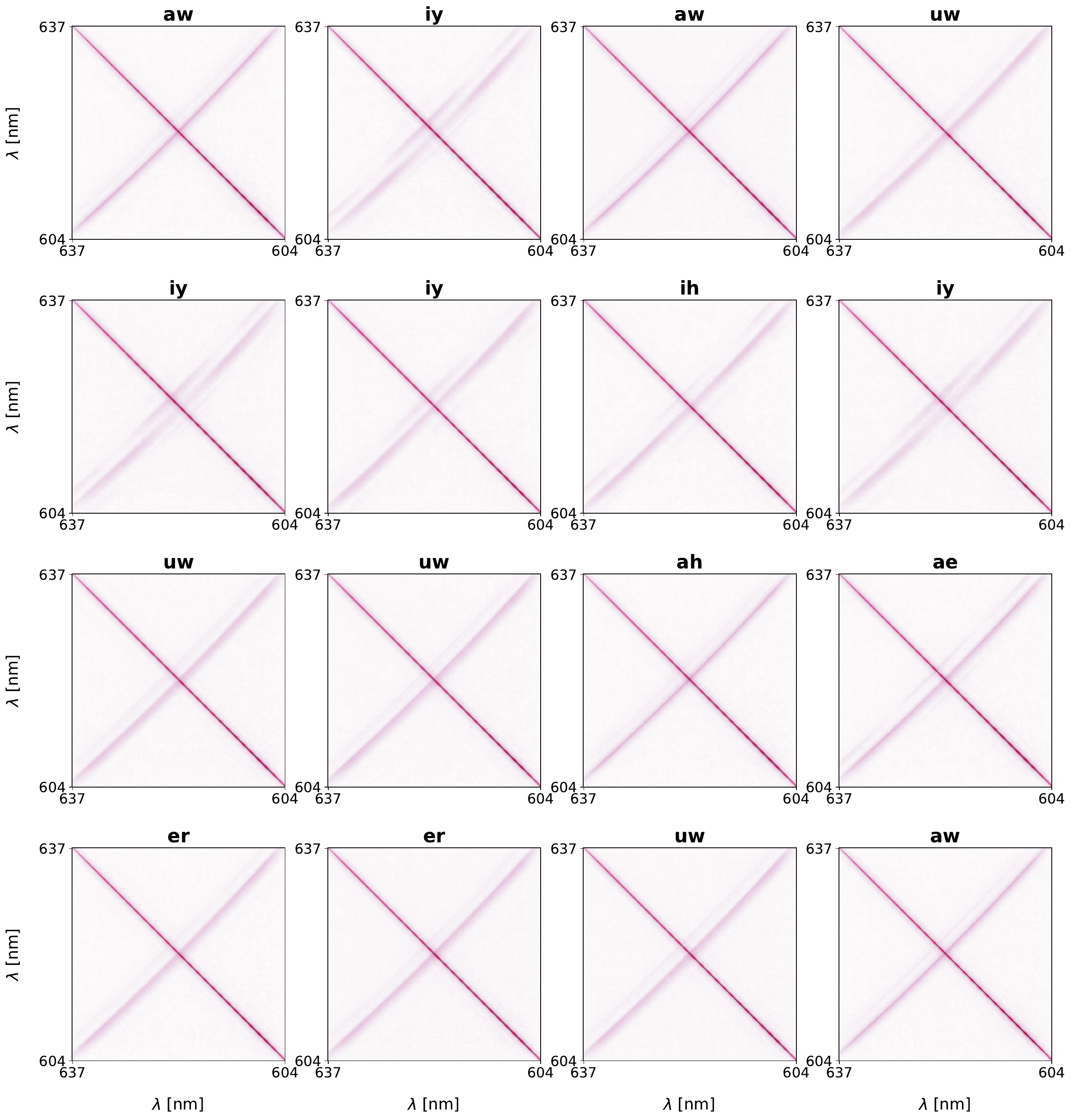}
	\caption{\textbf{Covariance matrices for different dataset samples for squeezed states}. The reconstructed covariance matrices are reported showing different correlation patterns for different AFC pump pulses. These pulses are directly linked to the dataset examples through the SLM phase encoding. Each matrix refers to different feature vectors corresponding to randomly selected vowel examples, as indicated in the title of each plot.}
	\label{fig6a}
\end{figure*}

\begin{figure*}[!htb]
	\includegraphics[width=0.99\textwidth]{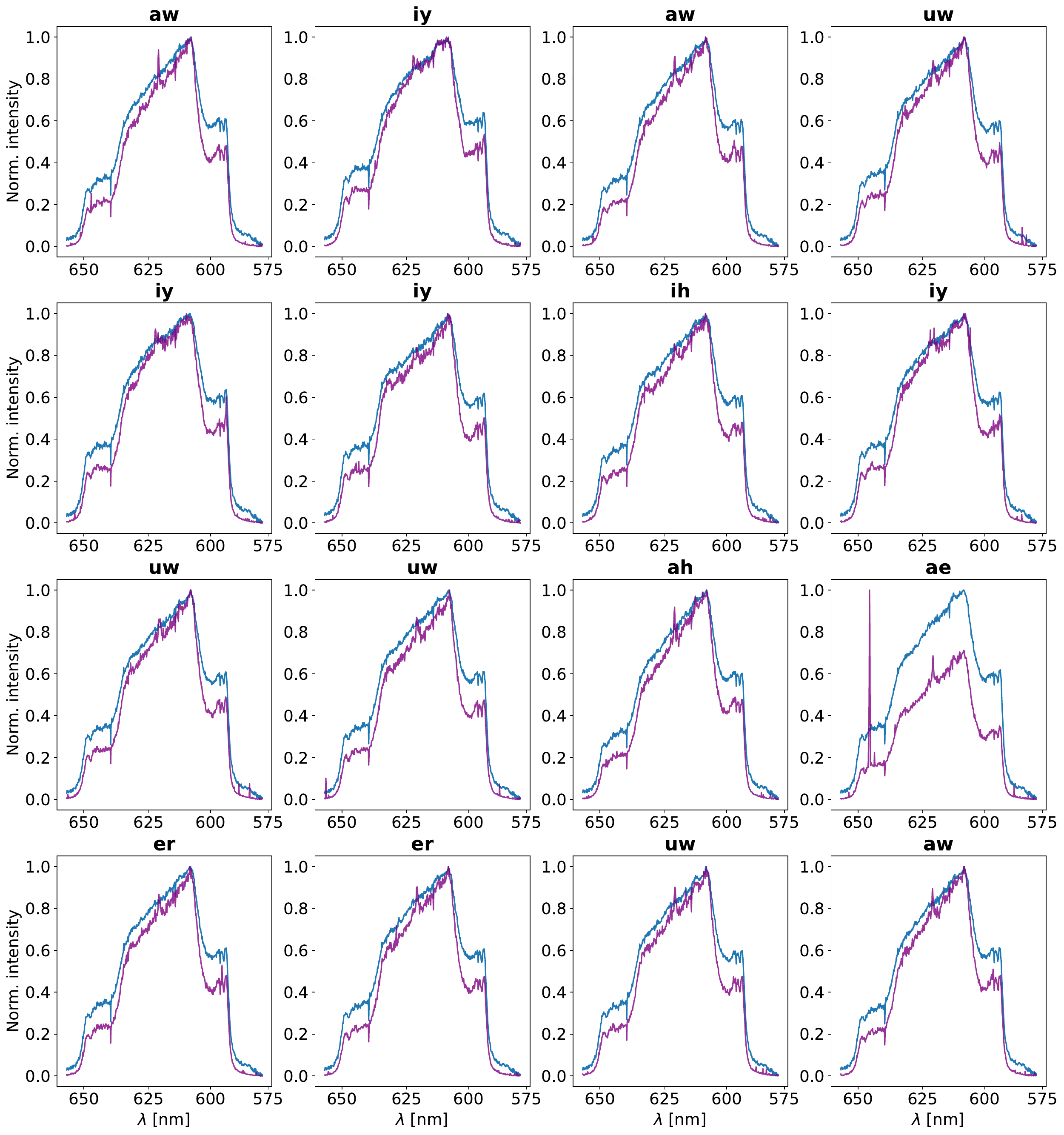}
	\caption{\textbf{Mean field vectors for different dataset samples for squeezed states}. The measured mean field vectors for randomly selected vowel examples, as indicated in the title of each plot, are presented. The blue line represents the normalized mean field, retrieved by averaging the counts across multiple camera frames. The purple line is the normalized diagonal of the reconstructed covariance matrix.}
	\label{fig7a}
\end{figure*}

For each classification task, we start by collecting multiple camera frames for each dataset sample, with a certain optical state. From these frames, we computed the mean field, represented as vectors of 512 elements, and the covariance matrix, which was flattened into vectors with $512^2$ elements. These vectors constituted the mean field training set and the covariance matrix training set, respectively. Once the data collection was completed, we trained a linear classifier as described in the main text. In Fig. \ref{fig6a}, we present examples of the reconstructed covariance matrices, and in Fig. \ref{fig7a}, we show the mean field obtained for random examples from the vowel dataset, with measurements taken using the quantum optical input.

\begin{figure*}[!htb]
	\includegraphics[width=0.99\textwidth]{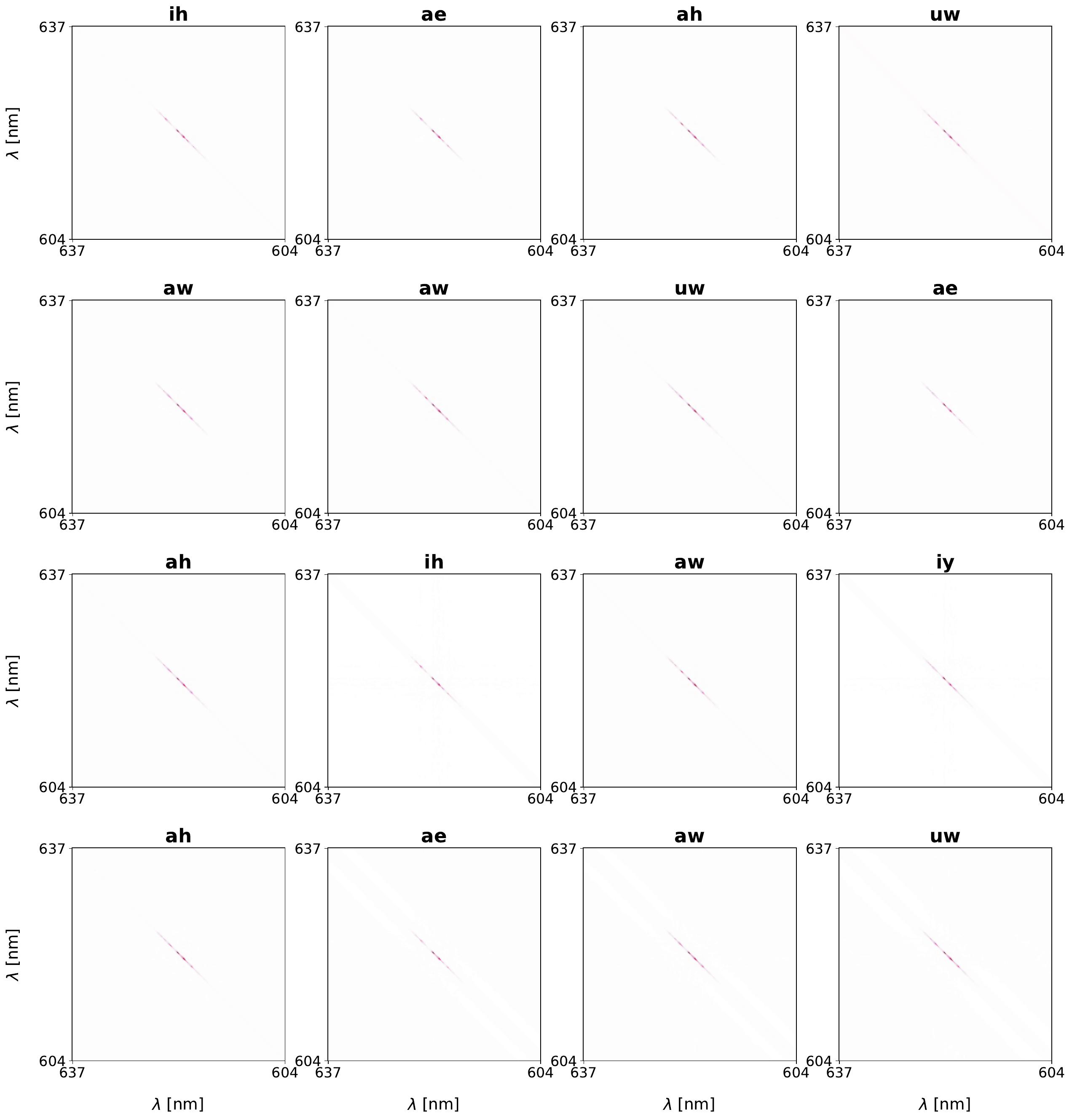}
	\caption{\textbf{Covariance matrices for different dataset samples for coherent states}. The reconstructed covariance matrices are reported when in the AFC process is involved the coherent input. Each matrix refers to different feature vectors corresponding to randomly selected vowel examples, as indicated in the title of each plot.}
	\label{fig6b}
\end{figure*}

\begin{figure*}[!htb]
	\includegraphics[width=0.99\textwidth]{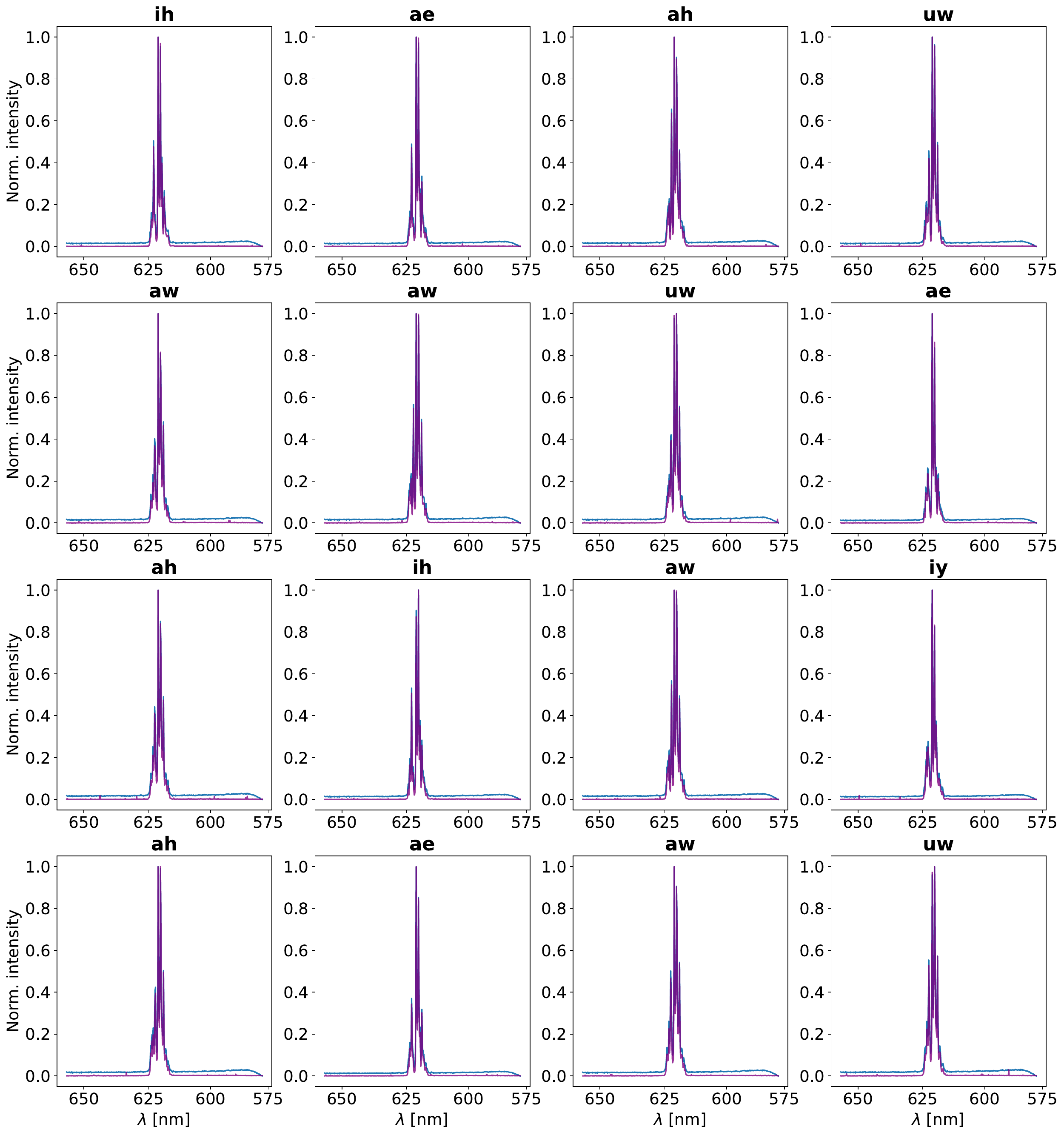}
	\caption{\textbf{Mean field vectors for different dataset samples for coherent states}. The measured mean field vectors for randomly selected vowel examples, as indicated in the title of each plot, are presented. The blue line represents the normalized mean field, retrieved by averaging the counts across multiple camera frames. The purple line is the normalized diagonal of the reconstructed covariance matrix.}
	\label{fig7b}
\end{figure*}

For comparison, we report the covariance matrices Fig. \ref{fig6b} and the mean field vectors Fig. \ref{fig7b} for the coherent state input.

The reconstructed covariance matrices have a high number of elements and, as evident in the plots, many elements of these matrices do not carry useful information. 
Given the relatively small size of the training dataset and the simplicity of the linear classifier, we aim to avoid overfitting. To mitigate this, we extract only the most informative elements from the covariance matrix. This is achieved by analyzing the variability between features across different classes and within the same class. Specifically, we employ the Analysis of Variance (ANOVA) F-test \cite{elssied2014novel}, which assesses how much the variance between the means of features across different classes exceeds the variance within each class.

\begin{figure*}[!htb]
	\includegraphics[width=0.99\textwidth]{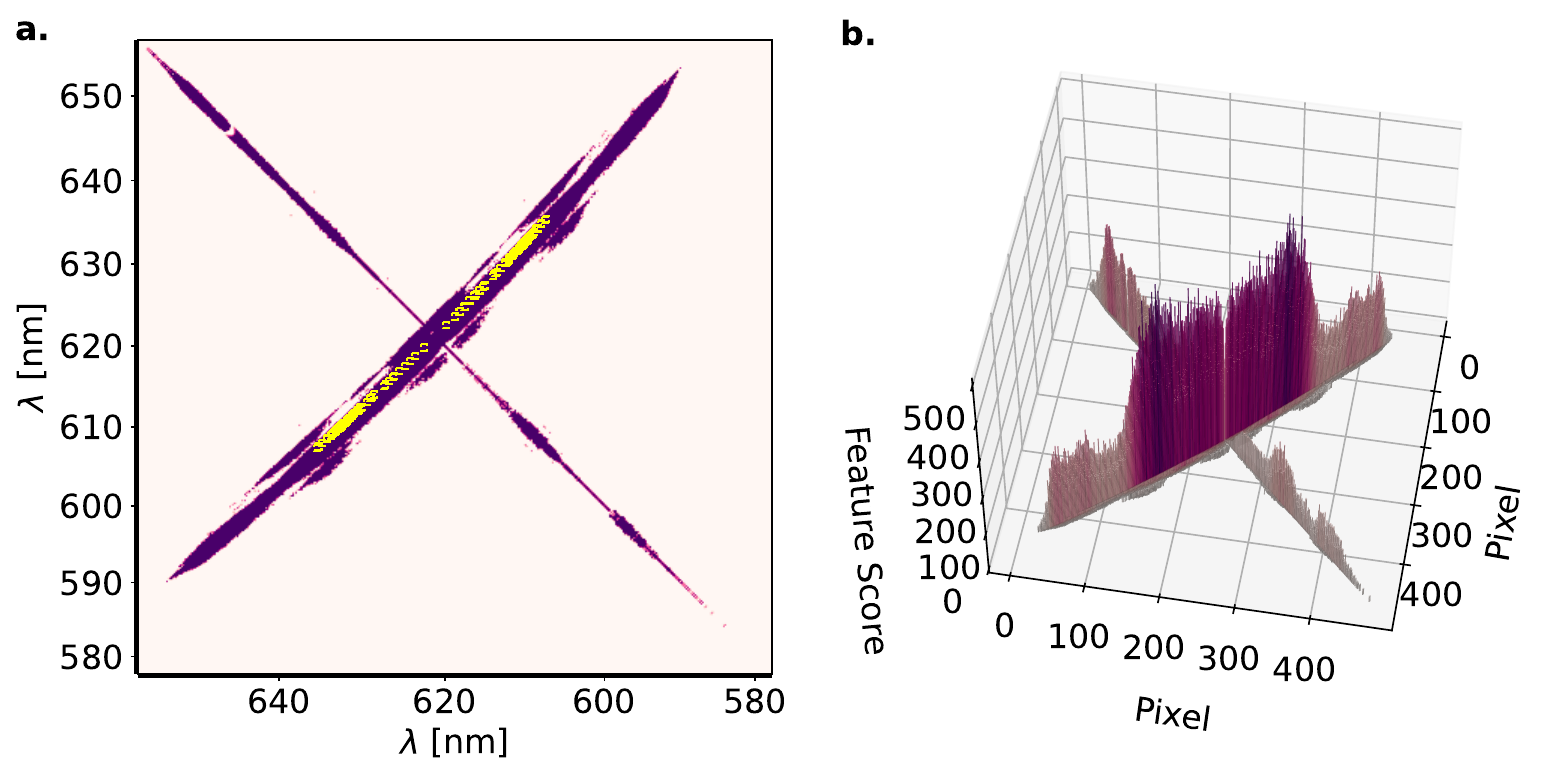}
	\caption{\textbf{Selected covariance features}. Panel a reports the most informative $15,000$ elements of the covariance matrix which are the ones used to train the linear classifier. The yellow squares highlight the elements with a feature score greater than $400$. In panel b are reported the F-values, representing the ratio of the variance between the average class features to the variance within each class. High F-values are related to features that are well differentiated between the classes.}
	\label{fig8a}
\end{figure*}

The F-test ranks the features based on the ratio of between-class variance to within-class variance, where a higher F-value indicates that the feature is strongly associated with the target class. Applying this feature selection method is particularly beneficial for high-dimensional covariance matrix datasets, as it reduces noise by eliminating irrelevant features and mitigates overfitting. We retained the top $15,000$ features, which correspond to the most informative regions of the covariance matrix, as shown in Fig. \ref{fig8a}.

The number of informative features is determined by selecting the value of k that yields the highest average classification accuracy on a validation set comprising $8\%$ of the dataset for all the inspected tasks. This cross-validation strategy provides a robust estimate of model performance and helps prevent overfitting by selecting only the most discriminative features.

\clearpage

\section{Optimization procedure}
\label{App_opt}

For each pair of classes with $m$ instances, the employed SVM solves the following convex optimization problem:
\begin{equation}
    \min_{w,b,\xi}\frac{1}{2} \mathbf{w}^T\mathbf{w} + \sum_{k=1}^m \xi^{(k)}, 
\end{equation}

such that 
\begin{equation}
   l^{(k)}(\mathbf{w}^T\mathbf{x}^{(k)}+b)\ge 1-\xi^{(k)} \text{ with } \xi^{(k)}\ge 0. 
\end{equation}
To maximize the margin that separates the different classes the quantity $||\mathbf{w}||=\mathbf{w}^T\mathbf{w}$ is minimized where $\mathbf{w}$ is the vector defining the orientation of the separating hyperplane, $b$ is a bias term, and $\xi$ is the distance for which some samples can be from the correct margin boundary.

To maximize the geometric margin between the classes, the norm of the weight vector $||\mathbf{w}||$ is minimized, since the margin is inversely proportional to it.

The equation in terms of Lagrangian multipliers can be rewritten as:
\begin{equation}
    \min_\alpha \frac{1}{2} \alpha^T Q \alpha - \sum_{i=i}^m \alpha_i,  
\end{equation}
where $Q$ is an $m\times m$ matrix whose elements are $Q_{kl} = l^{(k)}l^{(l)} K(x^{(k)},x^{(l)})$.

Since we employ a linear classifier the kernel function simplifies to the dot product of the feature vectors $K(x^{(k)},x^{(l)})=\mathbf{x}^{{(k)}^T}\mathbf{x}^{(l)}$, thus allowing the SVM to learn a linear decision boundary in the feature space.

Once the optimization problem is solved, the output for a new sample 
$\mathbf{x}$ is given by:
\begin{equation} 
f(\mathbf{x}) = \left( \sum_{i \in \text{SV}} \alpha_i l^{(i)} \mathbf{x}^{{(i)}^T} \right) \mathbf{x}^{(j)} + b, 
\end{equation}
where the sum runs over the support vectors only, i.e., the training samples for which $\alpha_i>0$. 

Therefore, the linear SVM defines the linear decision boundary that can be written in the form $\mathbf{w}^T \mathbf{x} +b$.

When considering multi-class problems, depending on the specific strategy chosen (\emph{one-vs-one} or \emph{one-vs-rest}), linear SVMs can lead to multiple binary classifiers that can be stacked into a matrix, recovering the generic linear classifier formulation
\begin{equation} 
f(\mathbf{x}) = \mathbf{W} \mathbf{x} + \mathbf{b}.  
\end{equation}

\end{document}